\documentclass{aa}  
\usepackage{float}

\usepackage{graphicx}
\usepackage{txfonts}
\usepackage[normalem]{ulem}
\usepackage{xparse}
\usepackage{expl3}

\usepackage{amsmath}
\usepackage{txfonts}
\usepackage{upgreek}
\usepackage[dvipsnames]{xcolor}
\usepackage{hyperref}
\usepackage[normalem]{ulem}

\hypersetup{
    colorlinks = true,
    linkcolor = blue,
    anchorcolor = black,
    citecolor = blue,
    filecolor = cyan,
    menucolor = red,
    runcolor = cyan,
    urlcolor = purple
}

\usepackage{diagbox}
\usepackage{siunitx}
\newcommand{\orcid}[1]{\href{https://orcid.org/#1}{\protect\includegraphics[width=8pt]{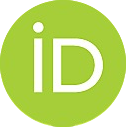}}}

\newcommand{\rev}[1]{\textcolor{black}{#1}} 

\begin{document}

   \title{Centrally concentrated star formation in young clusters}

    \author{
    Adilkhan Assilhan\orcid{0000-0001-6428-2084}\inst{1,2,3,4} \and
    Mordecai-Mark Mac Low\orcid{0000-0003-0064-4060}\inst{2,1} \and
    Brooke Polak\orcid{0000-0001-5972-137X}\inst{2} \and
    Ernazar Abdikamalov\orcid{0000-0001-5481-7727}\inst{5,6} \and
    Claude Cournoyer-Cloutier\orcid{0000-0002-6116-1014}\inst{7} \and
    Sean C. Lewis\orcid{0000-0003-4866-9136}\inst{8} \and
    Mukhagali Kalambay\orcid{0000-0002-0570-7270}\inst{4,9} \and
    Aigerim Otebay\orcid{0000-0003-3041-547X}\inst{9} \and
    Bekdaulet Shukirgaliyev\orcid{0000-0002-4601-7065}\inst{5,6,4}\thanks{Corresponding author.}
    }

    \institute{
    Department of Astronomy, Columbia University, 538 W 120th St., New York, NY, 10027, USA \and
    Department of Astrophysics, American Museum of Natural History, 200 Central Park W., New York, NY, 10024, USA \email{mordecai@amnh.org} \and
    Gumarbek Daukeyev Almaty University of Power Engineering and Telecommunications, 126/1 Baytursynuli St., Almaty, 050000, Kazakhstan \and
    Heriot-Watt University Aktobe Campus, K. Zhubanov Aktobe Regional University, 263 Zhubanov Brothers St., Aktobe, 030000, Kazakhstan \and
    Physics Department, Nazarbayev University, 53 Kabanbay Batyr Ave., Astana, 010000, Kazakhstan \email{bekdaulet.shukirgaliyev@nu.edu.kz} \and
    Energetic Cosmos Laboratory, Nazarbayev University, 53 Kabanbay Batyr Ave., Astana, 010000, Kazakhstan \and
    Department of Physics and Astronomy, McMaster University, 1280 Main St. W., Hamilton, ON, L8S 4M1, Canada \and
    Department of Physics, Drexel University, 3141 Chestnut St., Philadelphia, PA, 19104, USA \and
    Fesenkov Astrophysical Institute, 23 Observatory St., 050020 Almaty, Kazakhstan
    }

   \date{Received July XX, 2025; accepted month XX, 2025}

 
  \abstract
{
The study of star cluster evolution necessitates modeling how their density profiles develop from their natal gas distribution. Observational evidence indicates that many star clusters follow
 a Plummer{\rev{-like}} density profile.
However, most studies have focused on the phase after gas ejection, neglecting the influence of gas on early dynamical evolution. We investigate the development of star clusters forming within
 gas clouds, particularly those with a centrally concentrated gas profile. 
Simulations were conducted using the \texttt{Torch} framework, integrating the \texttt{FLASH} magnetohydrodynamics code into \texttt{AMUSE}. This permits detailed modeling of star formation, 
stellar evolution, stellar dynamics, radiative transfer, and gas magnetohydrodynamics. We study the collapse of centrally concentrated, turbulent spheres with a total mass 
of $2.5\times 10^3\, M_\odot$, investigating the effects of varying numerical resolution and star formation scenarios. The free-fall time is shorter at the center than at the edges of 
the cloud, with a minimum value of $0.55\,\mathrm{Myr}$. The key conclusions from this study are: (1) \rev{the final stellar density profile is more centrally concentrated than analytically 
predicted, reflecting the role of global gas collapse and feedback}; (2) sub-clusters can initially form even in centrally concentrated gas clouds; (3) gas collapses globally toward 
the center on the central free-fall time scale, contradicting the assumption in analytical models of local fragmentation and star formation; and
 (4) \rev{the mass of the most massive star formed is directly correlated with the cluster effective radius and inversely correlated with the velocity dispersion, 
 while the duration of star formation correlates with the star formation efficiency.}   
}


   \maketitle
%

\section{Introduction}
\label{sec:Intro}

Stars form within molecular clouds (MCs), in regions where gas accumulates under the influence of self-gravity \citep{2003ARA&A..41...57L}. Due to the critical role of this process, the formation of star clusters from MCs has been extensively studied via observations \citep{1978A&A....62..159B,2004ASPC..322..551O,2015ApJ...812..131K,2022A&A...662A...8M,2025jwst.prop.7534S} and simulations \citep{1984ApJ...285..141L,1997MNRAS.284..785G, 2006MNRAS.373..752G,2007MNRAS.380.1589B,2009ApJ...698.1659C,2012MNRAS.420.1503P,2018MNRAS.475.3511G,bek+2019,2023AAS...24110903L,Polak_2024a}. The formation and early evolution of star clusters 
involves not only the accumulation of gas but also the emergence of filamentary structures \citep{2004ARA&A..42..211E,2014MNRAS.444..336C,Alina+2019,2025AJ....169..203R}, as well as the influence of stellar feedback mechanisms, including stellar winds \citep{2013ApJ...770...25A,2014MNRAS.442..694D,Fierlinger+2016,2025arXiv250319745H}, radiation pressure \citep{Sales+2014,Rahner+2019,Kim+2018,2025AJ....169..133D}, and supernova explosions \citep{Hopkins+2012,2013MNRAS.431.1337R}. Observations of associated gas suggest that the upper limit of the cluster formation timescale is around 10 Myr \citep[e.g.][]{1989ApJS...70..731L}. Emerging clusters are often identified through the detection of H~\textsc{ii} regions \citep{2003IAUS..212..553H,2016A&AT...29..293G, Komesh+2024}.

While many stars form within embedded clusters, not all clusters survive their early evolution or remain gravitationally bound over long timescales. Observational studies indicate that a significant fraction of clusters disperse within 10--20~Myr, with less than 4--7\% of those initially embedded in MCs surviving the gas expulsion phase to remain bound \citep{2003ARA&A..41...57L,2007MNRAS.380.1589B}. However, clusters with a star formation efficiency (SFE) of 15--20\% or higher exhibit a greater probability of long-term survival \citep{1997MNRAS.284..785G,2011MNRAS.414.3036S,Bek+2017}. More recent studies suggest that, under specific initial conditions, clusters may remain bound with SFEs as low as 3--5\% \citep{Bek+2021}.

Whether a cluster remains bound may be due to the density profile of the MC from which it formed. \citet{2021MNRAS.502.6157C} suggest that rotating massive clusters, such as globular clusters, originate from centrally concentrated clouds in the early Universe. In particular, globular clusters, especially older and more massive ones, are known to exhibit strong central stellar concentrations \citep{2001ARA&A..39...99O}. This structure is often well described by the \citet{plummer1911} profile. Such centrally concentrated stellar profiles are also found in dense young clusters. For example, the Orion Nebula Cluster shows a stellar distribution within the central 0.5~pc that closely matches this profile \citep{2018MNRAS.473.4890S}. Beyond the Plummer radius $a$, the shape of the cluster appears to be influenced by the surrounding gas {filament}, illustrating the interplay between gas morphology and stellar dynamics during early cluster evolution.

\citet{2019MNRAS.487..364L} used the \texttt{AREPO} hydrodynamics code \citep{2010MNRAS.401..791S} to show that the evolution of MCs, driven by turbulent gas motions, leads to the formation of complex filamentary structures. Turbulent flows compress the gas under the influence of gravity, causing local collapse at filament intersections, where dense clumps form. This hub-filament structure leads to star formation \citep{hfs1-2009,hfs2-2024,hfs3-2024}. The resulting stellar sub-clusters often merge to form a larger stellar cluster. As stellar feedback increases, the surrounding gas either dissipates or collapses further, enhancing the filamentary structure. A subsequent study using \texttt{AREPO} showed that a gas cloud with a sufficiently centrally concentrated profile forms a single massive star cluster at its center, which then begins to accrete gas from the surrounding region \citep{2021MNRAS.502.6157C}. This accretion-driven formation results in compact central clusters.

\rev{The initial gas density profile and the resulting stellar distribution after gas expulsion are critical determinants of cluster survival. \citet{2000ApJ...542..964A} showed that a centrally concentrated stellar distribution substantially increases the likelihood of survival after gas removal. Building on this, \citet{2013A&A...549A.132P} developed a semi-analytical model in which the stellar density profile is steeper than the initial or residual gas profile. In their model, star formation proceeds through local gravitational collapse within the gas clump, with a constant star formation efficiency per free-fall time $\epsilon_{\mathrm{ff}}$, while the gas density profile remains fixed over the duration of star formation $t_{\mathrm{SF}}$. These assumptions naturally produce centrally concentrated stellar configurations, enhancing the long-term gravitational boundedness of the cluster.}

\citet{Bek+2017} tested the model of \citet{2013A&A...549A.132P}
by fitting the post-expulsion stellar distribution to a \citet{plummer1911} profile at different SFEs. Their approach assumed instantaneous gas expulsion and focused on whether the resulting stellar systems could remain gravitationally bound. They demonstrated that clusters can survive with a global SFE as low as 15\%, lowering the critical threshold from the previously suggested 20--30\% range \citep{2007MNRAS.380.1589B, 2011MNRAS.414.3036S,Brinkmann2016}. They argued that survival under such extreme, rapid gas-loss conditions implies an even greater likelihood of survival in more gradual, physically realistic scenarios. When evolved, their model clusters closely resemble observed open clusters in the solar neighbourhood \citep[][Weis et al. in prep.]{Kalambay2022,Abylay+2024, Marina+2025, Kalambay2025}.

Gas dynamics significantly affect early cluster evolution. \citet{2009ApJS..185..486P} and \citet{Bek+2017} showed that higher SFE at the moment of gas expulsion leads to a larger fraction of stars remaining gravitationally bound. Additionally, \citet{2011MNRAS.414.3036S} argued that the local stellar fraction within the half-mass radius is a more accurate predictor of cluster survivability than the global SFE. Gas removal alters the gravitational potential, often resulting in the expansion of surviving clusters by a factor of 3--4 \citep{2007MNRAS.380.1589B}. \citet{2010ApJ...710L.142F} further showed that the SFE is closely linked to the surface density of protoclusters, highlighting the need to accurately model gas dynamics during formation.

Several recent studies, such as those by \citet{2019MNRAS.483.4999F} and \citet{2023MNRAS.523.2083F}, have explored the effects of gradual star formation and gas dispersal using simplified hydrodynamic approaches within turbulent clumps. While these efforts provide valuable insight into the broader evolutionary trends of embedded clusters, our study extends their work by employing a state-of-the-art star formation framework that self-consistently couples gas dynamics, stellar feedback, and gravitational evolution at high resolution.

In our work, we extend the \citet{Bek+2017} model by beginning with pure gas and evolving the system during the star formation phase \emph{prior} to gas expulsion, incorporating the dynamical impact of directly modeled gas. We aim to quantify how the evolving gas density influences the cluster’s stellar distribution. 

We use the \texttt{Torch} framework \citep{wall2019,wall2020}, which combines magnetohydrodynamic (MHD) and N-body models within the Astrophysical MUltipurpose Software Environment \citep[\texttt{AMUSE};][]{2013A&A...557A..84P}. Modeling gas dynamics in star-forming regions requires accurately capturing turbulence, which plays a critical role in regulating gas fragmentation and collapse \citep{1981MNRAS.194..809L,2007ARA&A..45..565M}. Achieving a realistic treatment of turbulent gas structures, particularly the formation of dense clumps that initiate star formation, demands high spatial resolution, but this in turn requires significant computational resources. 

Previous studies, such as \citet{cournoyer-cloutier2023}, have employed the \texttt{Torch} framework to investigate the evolution of stellar clusters at high resolution. However, these simulations assumed a larger initial gas mass and included primordial binaries, which significantly increased computational costs and restricted the exploration of a broader parameter space. \rev{Unlike \citet{cournoyer-cloutier2023}, we do not include primordial binaries and instead} adopt a more moderate initial gas mass and a simplified setup to allow systematic exploration across different initial conditions. 

In this work, we aim to model star cluster formation under controlled and computationally efficient conditions, while still capturing key aspects of gas dynamics and feedback processes.  We study how gas evolution influences the dynamical state and structural properties of stellar clusters formed within centrally concentrated environments.
\rev{
In our initial conditions, we implement 
the gas profile derived by \citet{Bek+2017}, enabling us to test whether the simulated clusters develop the expected Plummer-like structure after gas removal.}

This paper is organized as follows. In Sect.~\ref{sec:Method}, we describe the numerical methods and initial conditions. Sect.~\ref{sec:Results} presents the simulation results and discuss our findings in the context of previous studies. The limitations of our models are addressed in Sect.~\ref{sec:Model_limit}. Finally, we summarize our conclusions and outline directions for future work in Sect.~\ref{sec:Conclusion}.
\section{Method} 
\label{sec:Method}

We use the \texttt{Torch} framework \citep{wall2019,wall2020} to simulate the formation of star clusters from gas. \rev{In this work, we use \href{https://bitbucket.org/torch-sf/torch/src/torch-v1.0/}{\texttt{Torch version 1.0}}\footnote{\url{https://bitbucket.org/torch-sf/torch/src/torch-v1.0/}}.} \texttt{Torch} uses the \texttt{AMUSE} framework \citep{2009NewA...14..369P,2013A&A...557A..84P,2013CoPhC.184..456P,2019zndo...3260650P}, for which we use commit \texttt{044ca1f7a}. In the version used here, \texttt{AMUSE} connects the MHD code \texttt{FLASH} version 4.6.2 \citep{fryxell2000,dubey2014}, the stellar evolution code \texttt{SeBa} \citep{portegies-zwart1996,2012A&A...546A..70T}, the \texttt{SmallN} code \citep{Hut1995,2018araa.book.....P}, and the fourth-order Hermite N-body dynamics solver \texttt{ph4} \citep{mcmillan2012}. \rev{We use \texttt{Multiples} \citep{2018araa.book.....P} to regularize close binaries and few-body interactions that arise dynamically during the simulation. }

\texttt{FLASH} is an adaptive mesh refinement (AMR) grid code. We choose the \texttt{HLLD} Riemann solver \citep{2005JCoPh.208..315M} with reconstruction of the third-order piecewise parabolic method \citep{1984JCoPh..54..174C}. 

\subsection{\rev{Gravity Bridge}} 
\label{sec:bridge}

\rev{We couple the stellar and gaseous components using a variant of the \texttt{BRIDGE} method \citep{fujii2007, 2020CNSNS..8505240P}, which provides a Hamiltonian splitting framework to combine different integration schemes. In this approach, the interaction is advanced with a kick–drift–kick sequence: the stars receive a velocity update from the gravitational field of the gas, evolve under their internal $N$-body dynamics with \texttt{ph4}, and are then updated again by the gas. The gas undergoes an analogous sequence, receiving accelerations from the stars while being evolved hydrodynamically with \texttt{FLASH}. Although the bridge is formally symplectic, the use of a fourth-order Hermite scheme for stars and a second-order solver for gas introduces additional integration errors; however, these remain small for the timestep sizes considered \citep{wall2019}. Mutual forces are computed consistently by mapping stellar masses onto the AMR grid with a cloud-in-cell scheme and solving for the potential with the FLASH multigrid method \citep{ricker2008}, ensuring momentum conservation and symmetry in the coupling.}

\subsection{\rev{Stellar feedback}} 
\label{subsec:sfeedback}

\rev{The simulations include stellar feedback from photoionization, far ultraviolet (far-UV) radiation pressure and stellar winds, with sources restricted to stars of $M > 7\,M_\odot$. Stars with \( M \gtrsim 20\,M_{\odot} \) dominate feedback through ionizing radiation, winds, and supernovae \citep{2018NewAR..82....1H}, though more recent studies indicate that feedback effects may already become significant at somewhat lower masses, around $M \sim 13\,M_{\odot}$ \citep{2023A&A...670A.151Y}.}

\rev{Radiation transport is handled with the \texttt{FERVENT} module \citep{baczynski2015}, which follows the ray-tracing algorithm of \citet{wise2011} implemented in \texttt{ENZO}. Ionizing photon luminosities and mean photon energies are determined from stellar masses and ages: for stars with $T_* > 2.75 \times 10^4$ K these values are interpolated from the OSTAR2002 grid \citep{lanz2003}, while for cooler stars they are estimated from blackbody spectra \citep{stahler2004}. Heating from ionization is calculated as the difference between photon energies and the ionization potential of hydrogen. Far-UV photons in the range 5.6--13.6~eV are also included, which are absorbed by dust and eject photoelectrons. For lower-mass massive stars ($7 \lesssim M/M_\odot \lesssim 13$), the far-UV luminosity approaches that of ionizing radiation. Dust is restricted to gas cooler than $3 \times 10^6$ K to avoid unphysical cooling in regions where the grains would have been destroyed \citep{draine2011}. Stellar winds are implemented using the momentum injection scheme of \citet{wall2019,wall2020}, which is inspired by the inversion of the method described in \citet{simpson2015}. For further details on the feedback implementation, see \citet{wall2020}.}

\subsection{Mesh Refinement} 
\label{subsec:Refinement}

The computational domain has a side length of 13.75\,pc. We employ AMR, with a base grid $16^3$ at level 1, and a minimum refinement level of 2. The maximum refinement level, denoted by $n$, varies between simulations and takes values from 3 to 6 depending on the specific run. For brevity, we refer to the maximum refinement level as simply the "refinement level" throughout the paper. A detailed summary of the simulation parameters is provided in Sect.~\ref{InitialConditions}.

The AMR grid is refined based on temperature gradients and the local \citet{jeans1902} length
\begin{equation}
\lambda_j = [\pi c_s^2 / (G \rho)]^{1/2},
\end{equation}
where  \( c_s \) is sound speed, \(\rho\) is the gas density, and \( G \) is the gravitational constant. Refinement and derefinement are triggered when the Jeans length falls below 12 cell widths and above 24 cell widths, respectively. The temperature triggered refinement or derefinement occurs when the adapted \citet{Lohner1987CMAME..61..323L} estimator of temperature \rev{gradient} exceeds 0.98 or drops below 0.6, respectively.

At refinement level $n$, the maximum number of cells is given by $\mathrm{Cells}_{\mathrm{max}} = 16 \times 2^{n - 1}$. The corresponding minimum cell size $\Delta x_{\rm min}$ can be derived from $\mathrm{Cells}_{\mathrm{max}}$ and is listed in Table~\ref{tab:deluxesplit}. \rev{The minimum allowed resolution in the lower-density outer regions corresponds to $32^3$ grid cells, yielding a maximum cell size of $\Delta x_{\rm max} = 0.429\,\mathrm{pc}$.}

\begin{table}
\small
\caption{Numerical parameters and resolution settings for simulations.}\label{tab:deluxesplit}            
\centering                          
\begin{tabular}{llccccc}        
\hline\hline                 
{Model} & {$n$} & {Random} & {Cells$_{\rm max}$} & {$\Delta x_{\rm min}$} & {$r_{\rm sink}$} & {$\rho_{\rm sink}$} \\
{} & {} & {seed} & {} & {[pc]} & {[pc]} & {[cm$^{-3}$]}   \\ 
\hline                        
\texttt{n3s1-10} & 3 & 1--10 & $64^3$ & 0.214 & 0.64 & 1065.0 \\
\texttt{n4s1} & 4 & 1 & $128^3$ & 0.107 & 0.32 & 4262.0 \\
\texttt{n5s1} & 5 & 1 & $256^3$ & 0.053 & 0.16 & 17049.0 \\
\texttt{n6s1} & 6 & 1 & $512^3$ & 0.027 & 0.08 & 68198.0 \\
\hline                                   

\end{tabular}
\tablefoot{
Columns: (1) Simulations are named as \texttt{nXsY}, where  \texttt{X} indicates the refinement level and \texttt{Y} is the random see. For \texttt{n3s1-10}, 10 simulations are performed at $n=3$, each with a different random seed ranging from 1 to 10. 
(2) Refinement level. 
(3) Random seed number. 
(4) Maximum number of cells.
(5) Minimum cubical cell size.
(6) Sink accretion radius.
(7) Sink threshold density.
}
\end{table}

\subsection{Sink Properties and Star Formation} \label{subsec:Sink Properties and Star Formation}

Gas collapses on timescales of the free-fall time 
\begin{equation} \label{eq:ff}
t_{\rm ff} = [3\pi/ (32 G \rho)]^{1/2}.
\end{equation}
As the density increases, the free-fall time shortens, accelerating the collapse. This makes it challenging to resolve star formation in simulations due to the diminishing timesteps needed to resolve the collapse. To address this, we collect regions of dense gas into sink particles \citep{bate1995,krumholz2004} \rev{using the method implemented in FLASH by \citet{federrath2010}.} 

We set the sink threshold density \(\rho_{\rm sink}\) and accretion radius \(r_{\rm sink}\) such that, at the minimum cell size \(\Delta x_{\rm min}\), the Jeans length satisfies \(\lambda_J = 6 \Delta x_{\rm min} = 2 r_{\rm sink}\), which represents a modified version of the criterion used in \citet{1997ApJ...489L.179T}\rev{, consistent with the requirement of \citet{2001ApJ...547..280H} that, for high refinement levels, the local Jeans length be resolved by at least six grid cells to ensure adequate magnetohydrodynamic support.} When the density exceeds $\rho_{\rm sink}$, and the additional boundedness and convergence criteria  outlined in \citet{federrath2010} are met, a sink particle is formed with all the dense gas within $r_{\rm sink}$. \rev{After formation, the sink particle accretes all mass within its accretion radius that exceeds the threshold density.  Mass found within overlapping accretion zones accretes to the sink with the closest center of mass \citep{wall2019}. On each time step, each sink particle is moved to the center of mass of the stars and gas within its radius of accretion \citep{cournoyer-cloutier2021}. Sink particles exert gravitational force on gas, stars, and each other through direct summation. New sink particles cannot be created within the accretion radius of an existing sink, which prevents artificial overlap of sinks \citep{federrath2010}.}

Each sink particle \rev{on formation} is assigned a list of stellar masses randomly drawn using Poisson sampling \citep{2017MNRAS.466..407S,wall2019}, with masses sampled from the \citet{2002Sci...295...82K} initial mass function (IMF), ranging from $M_{\rm min} = 0.08\,M_\odot$ to $M_{\rm max} = 150\,M_\odot$. Each sink forms the next star in the list as soon as it accumulates enough mass. Thus, the total mass
\begin{equation}
M_{\rm total}=M_{\rm gas}+M_{\rm sink}+M_{\rm star}
\end{equation}
is conserved. \rev{Despite formation of primordial binaries being implemented in Torch \citep{cournoyer-cloutier2021}, no primordial binaries are initialized in these models. Other work by \citet{cournoyer-cloutier2021, cournoyer-cloutier2023,cournoyer-cloutier2024} investigates primordial binaries in detail. In particular, \citet{cournoyer-cloutier2023} shows that their presence does not affect the overall cluster structure (e.g., size and shape) while star formation is ongoing. In the following, massive stars are defined as stars with $M > 7\,M_{\odot}$.}

\subsection{Model Setup} 
\label{ModelSetup}

We adopt a star cluster model based on the centrally concentrated star formation scenario proposed by \citet{2013A&A...549A.132P}, in which star formation occurs locally at all positions within the gas clump with a constant efficiency per local free-fall time. This assumption results in a stellar distribution that forms self-consistently while maintaining a constant total density profile (stars plus gas) throughout the star formation process. 

\citet{Bek+2017} modified this model by requiring the final stellar distribution $\rho_\star(r)$ to follow a \citet{plummer1911} profile
\begin{equation}
\label{eq:plm}
\rho_\star(r) = \frac{3M_\star}{4\pi a_\star^3} \left(1 + \frac{r^2}{a_\star^2}\right)^{-5/2},
\end{equation}
where \(M_\star\) is the total stellar mass and \(a_\star\) is the Plummer radius. Based on the constraint of the constant total density profile
\begin{equation}
\label{eq:total_prfle}
\rho_0(r, t_{\mathrm{SF}}) = \rho_{\mathrm{gas}}(r, t_{\mathrm{SF}}) + \rho_{\star}(r),
\end{equation}
they recover the corresponding density profile of the residual (unprocessed) gas, ensuring consistency with the original formulation of  \citet{2013A&A...549A.132P}. 
The residual gas density profile \(\rho_{\rm gas}\) is then given by:
\begin{equation}
\label{eq:centrcon}
\rho_{\rm gas} = \frac{1}{k^2} - \frac{\rho_\star}{2} - \frac{1}{2}\left(K_2 + \frac{8}{k^6K_1}\right)^{1/2} + K_1,
\end{equation}
where \(k\), \(K_1\), and \(K_2\) are parameters determined by the constant SFE per free-fall time \(\epsilon_{\mathrm{ff}}\) and the star formation duration \(t_{\mathrm{SF}}\) (see Appendix~\ref{app:density} for further details). \rev{This form ensures consistency between the assumed Plummer distribution for the stellar component and the requirement that the sum of stars and gas reproduces the initial gas density profile throughout the star formation process.}

\begin{figure}[htbp]
    \centering
    \includegraphics[width=0.9\columnwidth]{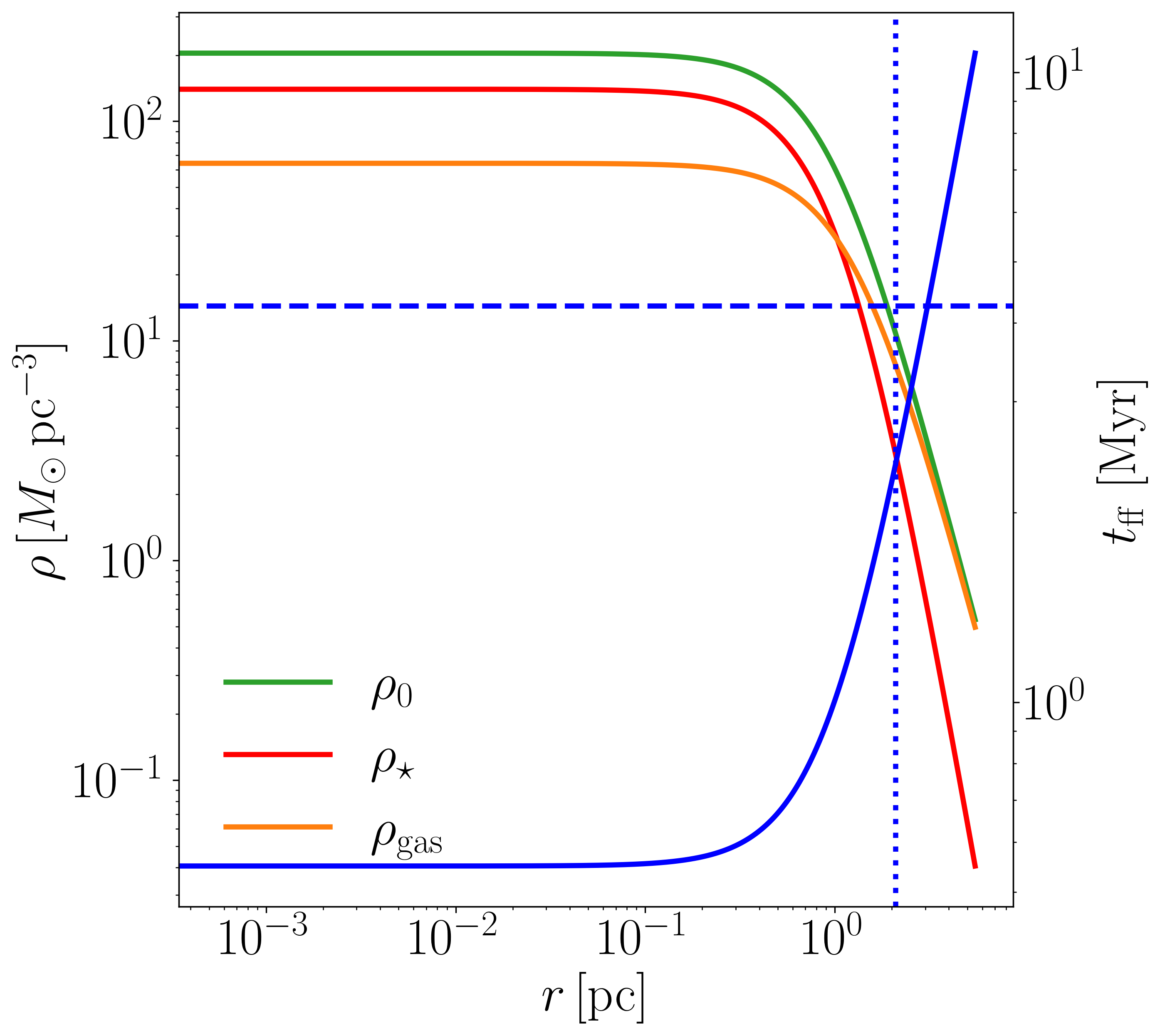}
    \caption{
    The initial density and free-fall time $t_{\rm ff}$ of the cluster derived by \citet{Bek+2017} as a function of radius. The initial gas density (Eq.~\ref{eq:total_prfle}) is shown in \textit{green}, the final stellar density (Eq.~\ref{eq:plm}) in \textit{red}, and the final gas density (Eq.~\ref{eq:centrcon}) in \textit{orange}. The initial density profile is used directly in the simulation setup. 
    The free-fall time $t_{\rm ff}$ of the initial gas density distribution as a function of radius is shown in (\textit{solid blue}). The \textit{dashed blue} horizontal line indicates the free-fall time corresponding to the average density, while the \textit{dotted blue} vertical line marks the half-mass radius  $r_{\rm h}$ of the initial gas distribution. The local value of $t_{\rm ff}$ is shortest in the central regions and increases significantly toward larger radii.
    }
    \label{fig:prfle}
\end{figure}

We use the assumed total density $\rho_0$ (Eq.~\ref{eq:total_prfle}) as the initial gas density profile for our simulations, as shown in  Figure~\ref{fig:prfle}.  This figure also shows the final gas (Eq.~\ref{eq:centrcon}) and stellar (Eq.~\ref{eq:plm}) density profiles derived by \citet{Bek+2017} from their assumption of a final Plummer profile, together with the local free-fall time $t_{\rm ff}$ of the initial gas distribution. The stellar component then forms self-consistently during the simulation. 

The initial free-fall time $t_{\text{ff}}(r)$ of the gas distribution provides insight into the early dynamical evolution of the system. As shown in Figure~\ref{fig:prfle}, the free-fall time is shorter in the central regions and increases substantially at larger radii, indicating that the gas in the central core collapses much more rapidly than the gas in the outer regions. The minimum free-fall time in the center is $t_{\text{ff}} = 0.55$ Myr. We evolve the system for at least $4.5\,t_{\text{ff}}$ with different resolutions, ensuring that the gas within the half-mass radius fully collapses. The free-fall time at the gas half-mass radius is about $2.39$ Myr initially.

\begin{table}
\caption{Initial physical parameters adopted in the simulations.}
\label{tab:Init1}
\centering
\small
\begin{tabular}{lc}
\hline\hline
Parameter & Value \\
\hline
Gas mass, $M_{\rm gas}$ & $2513\,M_\odot$ \\
Sphere radius, $R_{\rm sphere}$ & $5.5$ pc \\
Plummer radius, $a_\star$ & $1.1$ pc \\
Minimum number of cells, Cells$_{\rm min}$ & $32^3$ \\
Maximum cell size, $\Delta x_{\rm max}$ & $0.429$ pc \\
Temperature, $T_{\rm cl} = T_{\rm amb}$ & $20$ K \\
Virial parameter, $\alpha_{\rm vir}$ & $0.5$ \\
Vertical magnetic field, $B_z$ & $3 \times 10^{-6}$ G \\
\hline
\end{tabular}
\end{table}

\subsection{Initial Conditions}
\label{InitialConditions}

\rev{We model the collapse of a spherical turbulent gas cloud with a centrally concentrated density profile, embedded in a uniform background medium, and threaded by a weak uniform vertical magnetic field.} The cloud has a radius $R_{\rm sphere} = 5.5\,\mathrm{pc} = 5 a_\star$. \rev{To choose the cloud density profile, we use} an analytic \rev{model from \citet{Bek+2017} with a} stellar mass of $M_\star = 827\,M_\odot$ \rev{and 25\% SFE, but rescale it to the density profile it had prior to star formation when it was pure gas.  This ensures} 

comparability to the structural properties adopted in previous studies \citep{Bek+2017,bek+2018,bek+2019,2020IAUS..351..507S}.

The initial cloud gas mass is $M_{\rm gas} = 2513\,M_\odot$.
Our computational box is slightly larger than the spherical cloud of size $5a_\star$ that we initialize. Beyond the cloud ($r > 5.5\,\mathrm{pc}$), we initialize a uniform, initially isothermal background gas at the same temperature $T_{\rm amb} = 20 K$ and density as the surface of the cloud, \rev{with zero velocity, to ensure a smooth pressure transition at the boundary and avoid numerical artifacts.}
For comparison, \citet{Bek+2017} measure the global SFE within $r < 10a_\star$, as their models exhibit a centrally peaked SFE profile \citep[see Fig.~2 in][]{BekPhDT}. Within $10a_\star$, the centrally concentrated stellar profile, \rev{which follows approximately $\rho_\star \propto r^{-3.3}$, encloses a total mass (stars + gas) of $3260\,M_\odot$.}
\rev{Truncating the cloud at $5.5\,\mathrm{pc}$ and embedding it in this uniform medium increases the total gas mass in the simulation domain to $3597\,M_\odot$, $\sim10$\% more than the total mass of gas and stars used in the model by \citet{Bek+2017}. This approximation has minimal impact on the cloud’s centrally concentrated dynamics.}
The adopted physical parameters of the cloud are summarized in Table~\ref{tab:Init1}. 

\rev{Within $r \leq R_{\rm sphere} = 5.5,\mathrm{pc}$, the cloud is initialized with a turbulent velocity field generated \citep{2015HiA....16..614W} following a \citet{1941DoSSR..30..301K} velocity spectrum. The surrounding background gas ($r > R_{\rm sphere}$) has zero velocity to initialize a static, uniform medium. We set the integrated turbulent virial parameter of the cloud to $\alpha_{\rm vir} = 0.5$ by setting the cloud's total turbulent kinetic energy to be half of the total gravitational potential energy. The initial temperature of 20 K defines the thermal internal energy of the gas and the corresponding sound speed, which is independent of the turbulent kinetic energy. Thus, thermal pressure support and turbulent support enter the initial conditions as separate contributions. This approach does not fully balance energies in a centrally concentrated cloud, as higher central density would require stronger turbulent support. However, dissipative processes such as radiative cooling in interstellar clouds anyway prevent sustained virial equilibrium \citep[e.g.][]{2016ApJ...824...41I, 2017ApJ...850...62I, 2019MNRAS.482.2697S}.}

\rev{We use outflow boundary conditions to allow gas to exit the computational domain, while permitting inflow from boundary regions if the flow is directed inward, preventing unphysical vacuum formation at the edges.}

\subsection{Models}
\label{subsec:models}
We performed a series of simulations at different refinement levels to explore the influence of numerical resolution and stochastic IMF sampling on cluster formation. The corresponding numerical parameters for each simulation are listed in Table~\ref{tab:deluxesplit}. The refinement level $n$ was varied between $n = 3$ and $n = 6$, with maximum effective resolutions ranging from $64^3$ to $512^3$ cells. The minimum cell size $\Delta x_{\rm min}$, sink accretion radius $r_{\rm sink}$, and sink formation threshold $\rho_{\rm sink}$ were adjusted accordingly to ensure that the local Jeans length is properly resolved. For the lowest resolution case ($n=3$), we conducted ten independent realizations with different random seeds to capture statistical variations resulting from stochastic IMF sampling. The simulation naming convention is \texttt{nXsY}, where \texttt{X} denotes the refinement level and \texttt{Y} specifies the random seed.

\section{Results} 
\label{sec:Results}

We now examine what our numerical results reveal about the ultimate structure of the star clusters formed in our simulations.

\subsection{Cluster Structure}
\label{sec:plummer_fit}

We here compare the density profiles and morphologies of our clusters with the analytically predicted, spherically symmetric, Plummer density profile \(\rho_\star(r)\) described by Equation~(\ref{eq:plm}), with a Plummer radius of $a_\star = 1.1$~pc \rev{and a total stellar mass of \(M_\star = 780\,M_\odot\) (see Fig.~\ref{fig:prfle}). We also compare our simulated stellar density profiles with the more general Elson-Fall-Freeman (EFF) profile \citep{Elson1987}, which the Plummer profile is a special case of.
Finally, we compare our results with observational data to evaluate their consistency with observed star cluster properties.}

\rev{The three-dimensional EFF density distribution is given by \citep{Jeremy2022}}
\begin{equation}
\label{eq:EFF} 
\rev{\rho_{\rm eff}(r) = \rho_0 \left(1 + \frac{r^2}{a_{\rm eff}^2}\right)^{-(\gamma_{\rm eff}+1)/2},}
\end{equation}
\rev{where $\rho_0$ is the central density, $a_{\rm eff}$ is the scale radius and $\gamma_{\rm eff}$ determines the outer slope of the density profile. The Plummer profile corresponds to the special case $\gamma_{\rm eff} = 4$, while the EFF form allows $\gamma_{\rm eff}$ to vary. We therefore fit both a Plummer profile, with Plummer mass $M_{\rm pl}$ and scale raduis $a_{\rm pl}$ as free parameters, and an EFF profile, with $\rho_0$, $a_{\rm eff}$, and $\gamma_{\rm eff}$ as free parameters, to the simulated stellar distributions to test whether the clusters are consistent with the analytically predicted Plummer profile.}

\rev{The stellar density profiles of simulated clusters are constructed by binning stellar positions in 20 logarithmically spaced radial shells between the minimum and maximum stellar radii. In each shell, the mean density}
\begin{equation}
\label{eq:iter} 
\rev{\rho_i = \frac{\sum_j m_j}{\tfrac{4}{3}\pi\left(r_{i+1}^3 - r_i^3\right)},}
\end{equation}
\rev{where the sum extends over all stars with radii \(r_i \leq r_j < r_{i+1}\). Empty bins are excluded from the analysis. The cluster center is determined iteratively by recalculating the mass-weighted centroid while the enclosing radius is reduced by 10\% at each step, until fewer than 32 stars remain. This shrinking-sphere procedure yields a robust definition of the cluster center while naturally converging toward the densest subcluster \citep[see e.g.][]{Bek+2017,shukirgaliyev2019violent,bek+2019,ussipov2024fractal,Abylay+2024,bissekenov2024exploring, Abylay+2025, Marina+2025}.}

\begin{figure*}[htbp]
    \centering
    \includegraphics[width=0.88\textwidth]{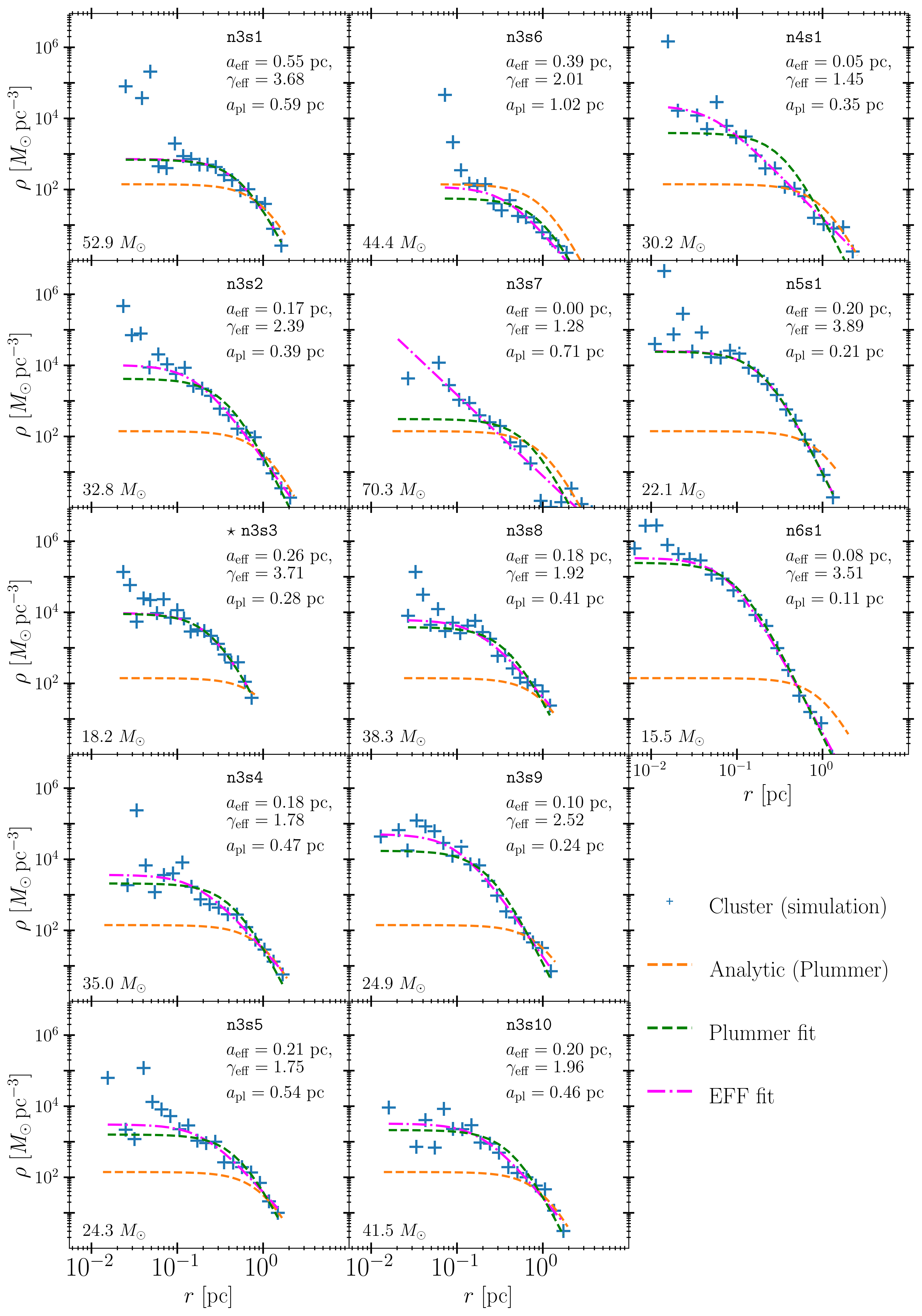}
    \caption{\rev{
    Stellar density profiles at \(4\,t_{\rm ff}\) (\emph{blue points}) are compared with the analytically predicted Plummer profile for the stellar distribution (\emph{dashed orange line}), as well as best-fit Plummer (\emph{dashed green}) and EFF (\emph{dash-dotted magenta}) models.
    The mass of the most massive star formed in each simulation is indicated in the lower-left corner of each panel. The panel marked with a $\star$ (\texttt{n3s3}) is shown at a slightly earlier time; see also the text for details on sub-cluster cases.
    }
    }\label{fig:Prfle_b}
\end{figure*}

\rev{Figure~\ref{fig:Prfle_b} presents the stellar density profiles of the simulated clusters \emph{(blue points)} at \(4\,t_{\rm ff}\). The cluster is centered on the densest region, and only the stars within the radius enclosing 95\% of the total stellar mass are used for calculation of the density profile. We compare them with the analytically predicted Plummer profile \emph{(dashed orange line)} for stellar clusters (shown in Figure~\ref{fig:prfle} and given by Eq.~\ref{eq:plm}), as well as with the best-fit Plummer \emph{(dashed green line)} and EFF \emph{(dash-dotted magenta line)} profiles. The fits are obtained with the \texttt{curve\_fit} routine from \texttt{scipy} \citep{2020NatMe..17..261V}, which applies the Levenberg–Marquardt algorithm \citep{Levenberg1944} for nonlinear least squares optimization. To avoid artificial central peaks caused by a few massive stars, the innermost region (\(0.01-0.2\,\mathrm{pc}\), depending on the simulation) is excluded from the fits, and only stars within the radius enclosing 95\% of the total stellar mass are used.} 

\rev{This comparison highlights the differences between the evolved stellar systems and the predicted analytic distribution and shows how the Plummer and EFF models capture the structural properties of the clusters. The fitted Plummer and EFF scale radii provide a quantitative measure of the cluster compactness. Some simulations agree well with a Plummer profile (e.g., \texttt{n3s1}, \texttt{n3s3}, \texttt{n5s1}, and \texttt{n6s1}), while others deviate, as indicated by the slope of the density tail in the EFF profile. Nevertheless, the overall structure is reasonably well described by both profiles. With increasing refinement level, the stellar density profiles become smoother and more centrally concentrated, as reflected by the higher densities within \(0.1\,\mathrm{pc}\) (e.g., comparing \texttt{n3s1}, \texttt{n4s1}, \texttt{n5s1}, and \texttt{n6s1}). Although the analytically predicted Plummer radius is \(a_\star = 1.1\,\mathrm{pc}\), the fitted scale radii are systematically smaller, both in \(a_{\rm pl}\) and in \(a_{\rm eff}\). This shows that the star clusters formed in our models are more compact than predicted.  Thus, a centrally concentrated gas distribution under realistic conditions leads to the formation of a more compact stellar cluster than predicted by the analytic model, which assumed local gravitational collapse. }

\rev{In model \texttt{n3s3}, the snapshot marked by a $\star$ is shown 40~kyr earlier than \(4\,t_{\rm ff}\), because a very massive star of \(60\,M_\odot\) forms shortly after this time. At the displayed epoch, however, the dominant feedback arises from an \(18\,M_\odot\) star, which controls the gas evolution at that stage. In model \texttt{n3s7}, a \(70\,M_\odot\) star forms and the system develops a clear sub-cluster structure, which prevents meaningful Plummer or EFF fitting. We also note that models \texttt{n3s6} and \texttt{n4s1} form sub-clusters. In cases of sub-clusters, the profiles are centred on the densest sub-cluster, so the outer regions may not be fitted correctly.}

This conclusion aligns with observations of young clusters such as the Orion Nebula Cluster (ONC). \citet{2018MNRAS.473.4890S} \rev{derived the stellar mass distribution in the Orion Nebula Cluster primarily from Spitzer-identified YSOs \citep{2016AJ....151....5M}, augmented with COUP X-ray data in regions of high nebulosity \citep{2005ApJS..160..379F}.  They found that the central cluster core ($\lesssim 0.7$ pc) contains \(\sim 1000\)--\(1300\,M_{\odot}\), surrounded by a more diffuse halo extending to $\sim 2$ pc (see their Fig.~4). The inner structure is well fit by a Plummer model with $a_\star \sim 0.36$~pc.}

\subsection{Sub-cluster Formation}
\begin{figure*}[!htbp] 
    \centering
    \includegraphics[width=0.9\textwidth]{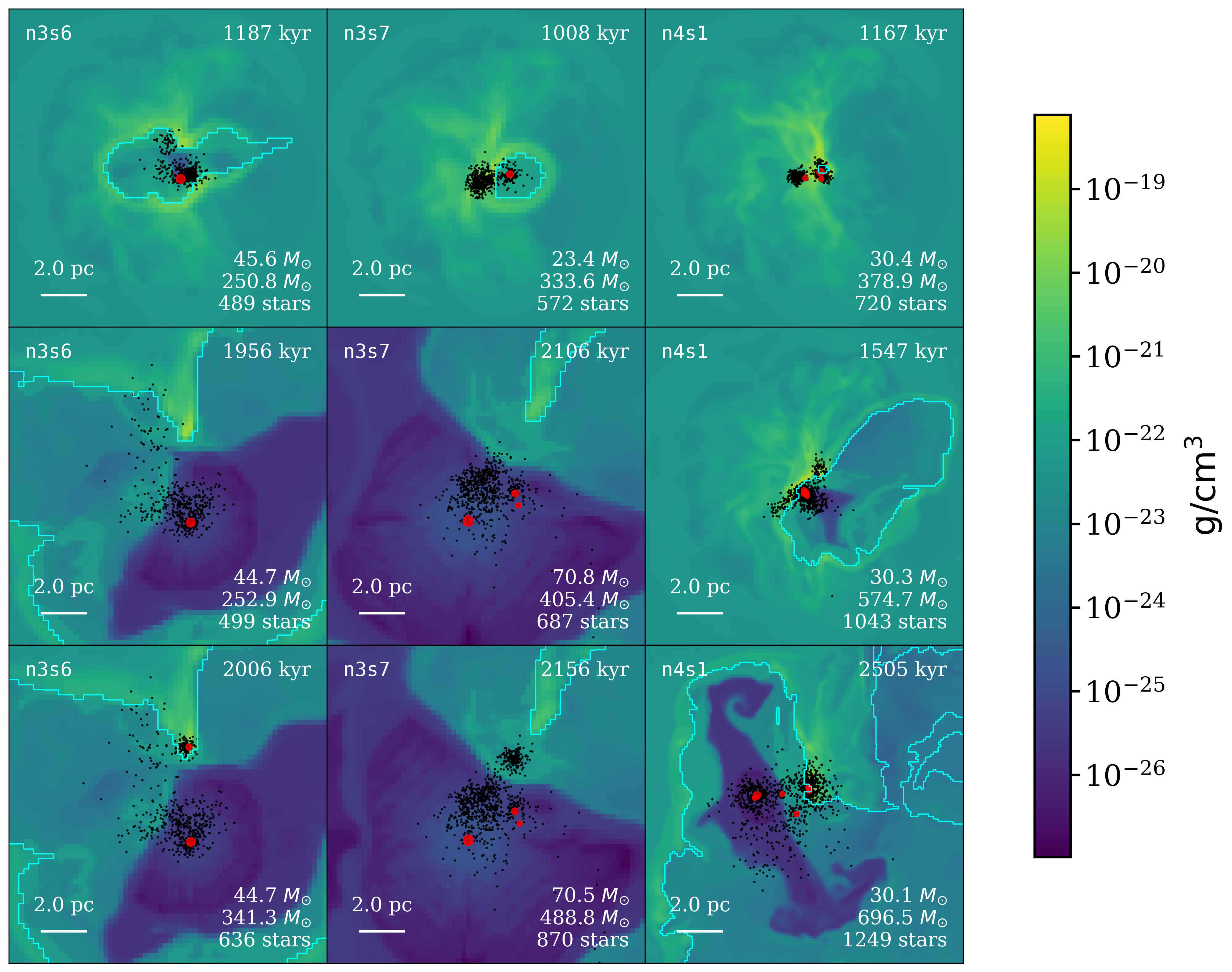}
    \caption{Gas density slices and stellar distribution projections from the \texttt{n3s6}, \texttt{n3s7}, and \texttt{n4s1} models. In these models sub-cluster formation occurs. For \texttt{n3s6} and \texttt{n3s7}: {\em (top row)} onset of stellar feedback; {\em (middle row)} prior to the formation of the second cluster; {\em (bottom row)} after the formation of the second cluster. For \texttt{n4s1}: {\em (top row)} two clusters are present; {\em (middle row)} clusters merge into a single structure; {\em (bottom row)} the system separates again into two clusters. Ionization fronts are indicated by the cyan line. {\em Black dots} indicate individual stars, with massive stars shown in {\em red}. The symbol size of massive stars is proportional to their mass. Annotations in each panel include: {\em (bottom left)} scale bar; {\em (bottom right)} mass of the most massive star, total stellar mass within 5.5~pc, and total number of stars; {\em (top left)}  simulation label; {\em (top right)} simulation time. The color scale represents the gas density in g cm\(^{-3}\).}
    \label{fig:Fbound}
\end{figure*}

Figure~\ref{fig:Fbound} shows the stellar distribution and gas density evolution for simulations \texttt{n3s6}, \texttt{n3s7}, and \texttt{n4s1}, which were chosen because they form multiple stellar sub-clusters. In all cases, the formation of two distinct sub-clusters is clearly observed. In simulation \texttt{n4s1}, the dynamical interaction between these sub-clusters becomes particularly evident: initially, the two sub-clusters form in close proximity and exhibit elongated structures (top row), subsequently merge into a single system (middle row), and later separate again into two sub-clusters (bottom row). 
In contrast, in \texttt{n3s6} and \texttt{n3s7}, the first sub-cluster forms early around the first massive star (top row), and stellar feedback then disperses the surrounding gas and shifts the dense region spatially (middle row), enabling the later formation of a second sub-cluster in the displaced gas (bottom row). 

These results demonstrate that even in initially centrally concentrated gas clouds, multiple sub-clusters can emerge depending on the timing and location of massive star formation. This extends the results of \citet{cournoyer-cloutier2023} and \citet{Polak_2024a}, who found similar behavior in less centrally concentrated initial gas distributions. 
 
\rev{Notably, the formation of multiple stellar sub-clusters is relatively uncommon in our study: only 3 out of 13 simulations (spanning various refinement levels and random seeds) exhibit clear sub-cluster formation driven by massive star feedback, whereas the majority evolve into single, merged systems sustained by prolonged central accretion. We identify two distinct pathways for sub-cluster formation in centrally concentrated clouds: (1) the sequential formation of a secondary stellar sub-cluster triggered by the displacement of dense gas regions due to feedback from an early massive star, as observed in simulations \texttt{n3s6} and \texttt{n3s7}; and (2) the near-simultaneous formation of two stellar sub-clusters in close proximity 
\texttt{n4s1}. In this case, feedback from massive stars disperses the gas, which reduces the system's binding energy and halts global collapse. As a result, when the sub-clusters fall toward the center, they become kinematically dominant and then separate again. This highlights the complex interplay between stellar dynamics, gas accretion, and feedback processes.}

\begin{figure*}[htbp]
    \centering
    \includegraphics[width=0.9\textwidth]{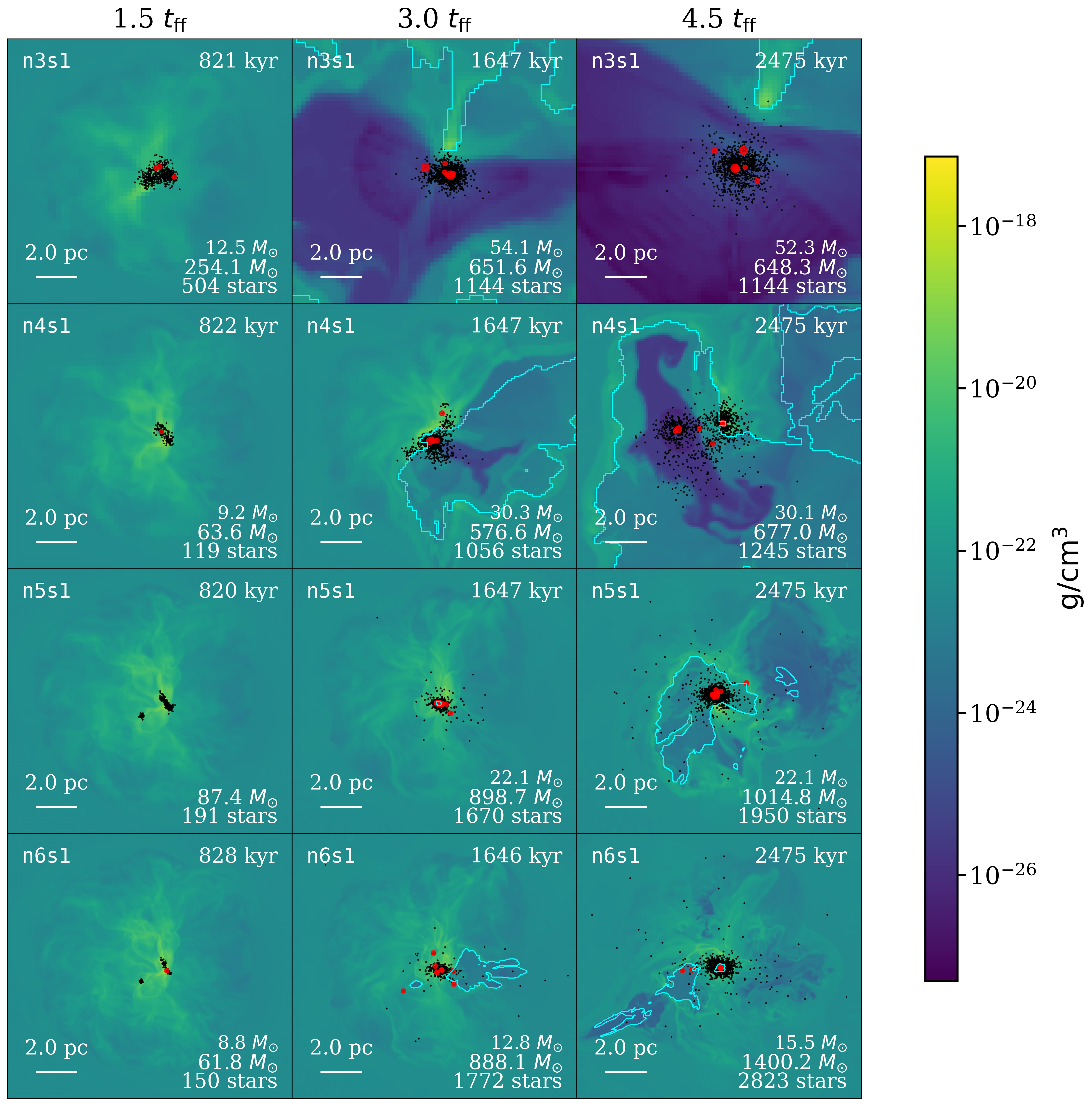}
    \caption{Images of gas density and stellar distribution at different times for different maximum grid resolutions (higher refinement level $n$ corresponds to finer grid resolution). Each row corresponds to a level; each column, to a simulation time ($1.5~t_{\rm ff}$, $3~t_{\rm ff}$, $4.5~t_{\rm ff}$). The ionization fronts are marked with cyan lines. Stars are black dots; massive stars are red, with sizes proportional to their masses. 
    In each panel, the notations are as follows: {\em (bottom left)} physical scale; {\em (bottom right)}  the mass of the most massive star, total stellar mass, and number of stars; {\em (top left)} model label; {\em (top right)} simulation time. The color scale represents the gas density in g~cm\(^{-3}\)}
    \label{fig:low}
\end{figure*}
\rev{At refinement levels \(n = 4\), 5, and 6, two star clusters form at the location of the peak density during the initial stages of the collapse. At $n=5$ and $n=6$, this occurs at time $1.5~t_{\rm ff}$ (see Fig.~\ref{fig:low}), and at $n=4$, about $\sim \,200$ kyr later. In the cases with \(n = 5\) and 6, the clusters merge (second column), forming a larger cluster that continues to accrete gas toward the center. In contrast, at $n=4$, the star clusters first merge and then split again into two clusters. This is probably due to the stronger feedback at $n \ = \ 4$, which is more intense than at higher refinement levels due to the formation of a more massive star. We conclude that star clusters can merge if gas continues accreting at the center for more than a dynamical time and the surrounding gas potential supports the merging process. These results are consistent with the simulation findings reported by \citet{2024ApJ...967...86K}.}

In summary, our analysis shows that different random initial conditions can lead to variations in the timing and location of massive star formation,
sometimes leading to the formation of multiple stellar sub-clusters.

\subsection{Impact of Massive Stars}
\label{subsec:ImpactMS}

In this subsection, we investigate the impact of the most massive stars on SFE and cluster structure. We quantify stellar feedback by measuring the expelled gas mass \( M_{\mathrm{loss,fb}} \) and examine the dependence of SFE on  \rev{the duration of star formation} \( t_{\mathrm{sf}} \).

\begin{figure}[!h]
    \centering
    \includegraphics[width=0.45\textwidth]{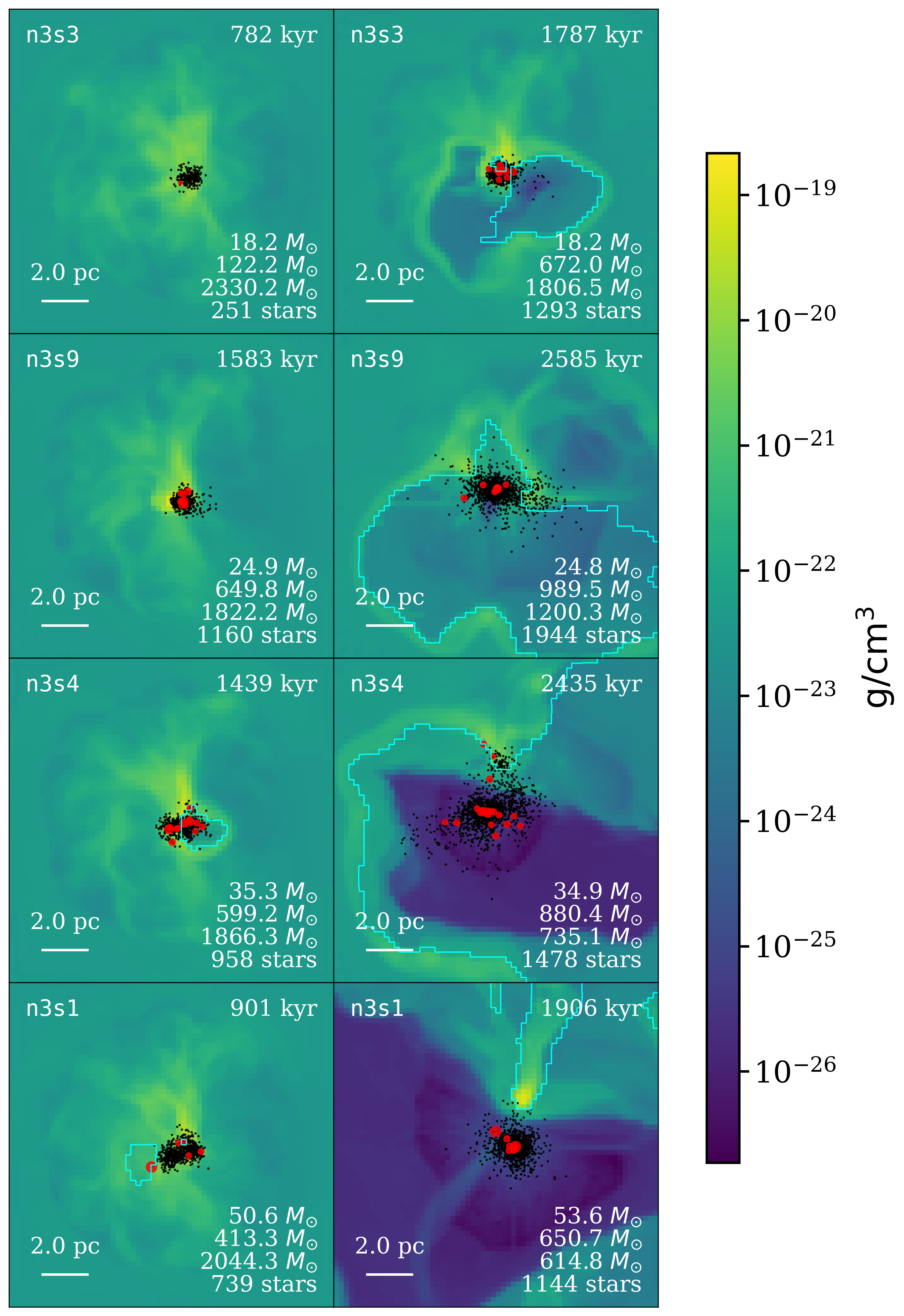}
    \caption{Slices of gas density and projected stellar distribution at the formation time in each model of the most massive star with \( M > 15\,M_{\odot} \) {\em (left column)}, and 1~Myr later {\em (right column)}, for the \texttt{n3s3}, \texttt{n3s9}, \texttt{n3s4}, and \texttt{n3s1} simulations. Ionization fronts are indicated by  {\em cyan} lines. {\em Black dots} represent stars, while massive stars are shown in {\em red}; massive star symbol sizes are proportional to stellar mass. In \texttt{n3s1}, two most massive stars form: one with \(50\,M_\odot\) at 891 kyr (triggering feedback), and a second with \(53\,M_\odot\) at 1187 kyr. Annotations in each panel are {\em (bottom left)} physical scale; {\em (bottom right)} mass of the most massive star, total stellar mass within 5.5~pc, total gas mass within 5.5~pc, and number of stars; {\em (top left)} model; {\em (top right)} simulation time. The color scale represents the gas density.
    }
    \label{fig:SFE0_25}
\end{figure}

\begin{figure}[htbp]
    \centering
    \includegraphics[width=0.45\textwidth]{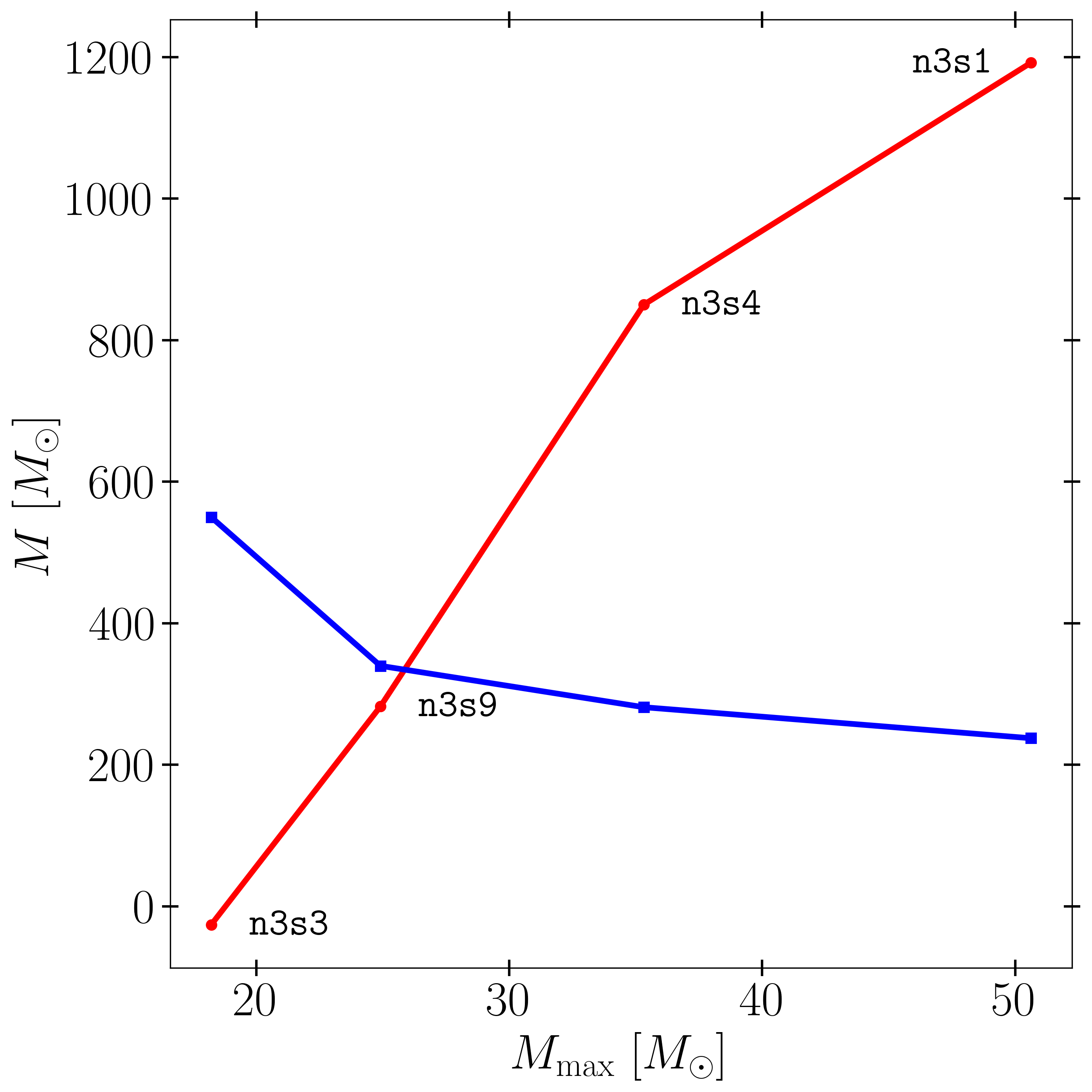}
    \caption{Stellar mass formed in first megayear after formation of most massive star {\em (blue)} and the gas mass lost due to feedback \( M_{\text{loss,fb}} \) {\em (red)} versus the mass of the most massive star for simulations \texttt{n3s3}, \texttt{n3s9}, \texttt{n3s4}, and \texttt{n3s1}. Simulation names are labeled next to each red dot. This is the same interval as in Figure~\ref{fig:SFE0_25}.}
    \label{fig:feedback}
\end{figure}

\begin{figure}[htbp]
    \centering
    \includegraphics[width=0.9\columnwidth]{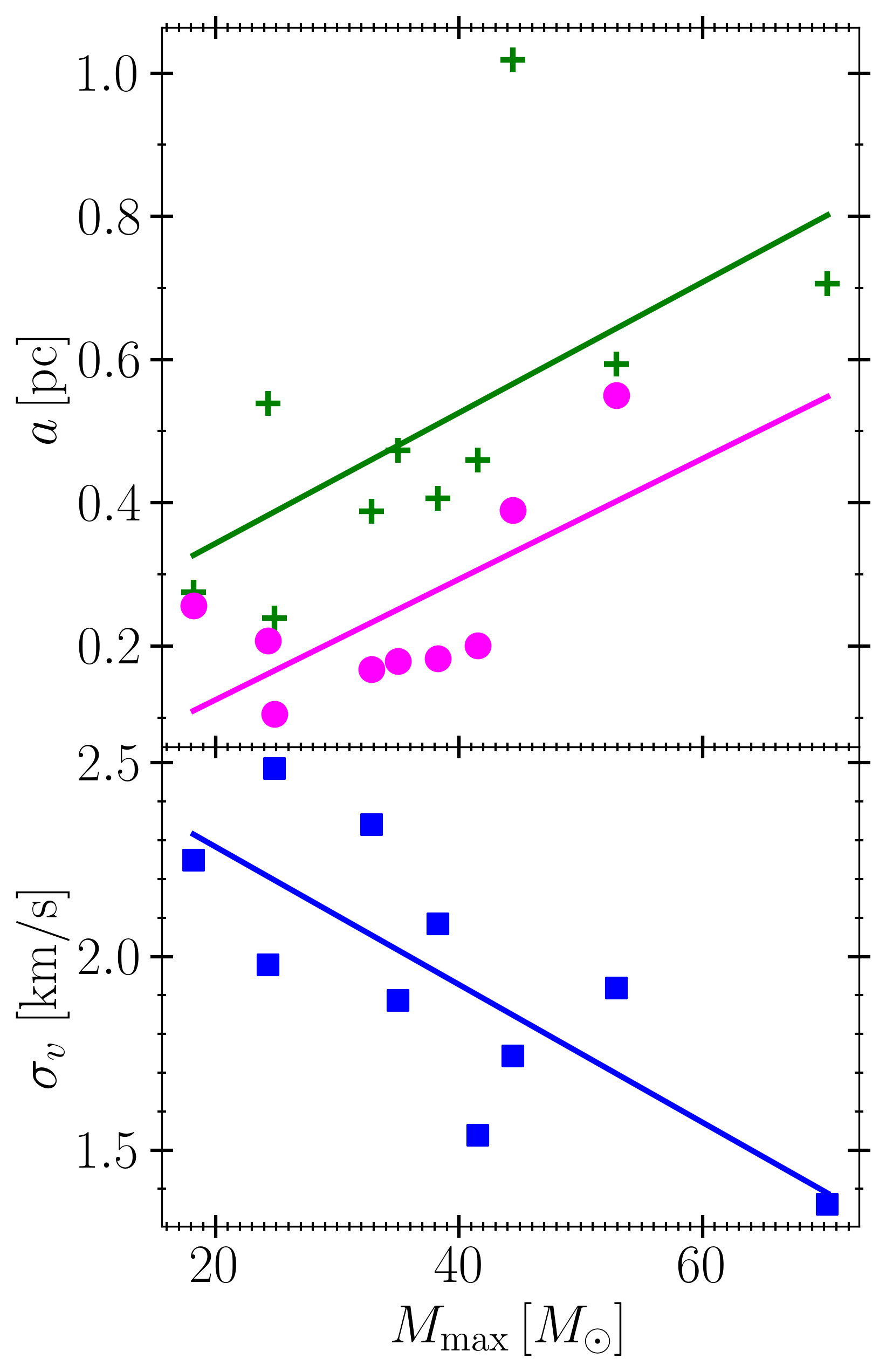}
    \caption{
    \rev{Relationship between the maximum stellar mass \( M_{\rm max} \), the Plummer radius \( a_{\rm pl} \), and the EFF scale radius \( a_{\rm eff} \) at \( 4t_{\rm ff} \), and the velocity dispersion \( \sigma_v \) for simulations at refinement level 3 with different random seeds. s. All quantities are measured using stars within the radius around the densest region of the cluster that encloses 95\% of the total stellar mass. The green line shows a linear fit to \( a_{\rm pl} \) (plus symbols), given by \( a_{\rm pl} / (1 \, \mathrm{pc}) = (0.0091 \pm 0.0041) [M_{\rm max} / (1 \, M_\odot)] + 0.1608 \). The magenta line shows a linear fit to \( a_{\rm eff} \) (circle symbols), given by \( a_{\rm eff} / (1 \, \mathrm{pc}) = (0.0084 \pm 0.0035) [M_{\rm max} / (1 \, M_\odot)] - 0.0437 \). The blue line shows a linear fit to \( \sigma_v \) (square symbols), given by \( \sigma_v / (1 \, \mathrm{km \, s^{-1}}) = (-0.0178 \pm 0.0051) [M_{\rm max} / (1 \, M_\odot)] + 2.6391 \). These linear fits are obtained with the \texttt{linregress} function from \texttt{scipy.stats}, which applies ordinary least-squares regression. The upper y-axis corresponds to \( a_{\rm pl} \) and \( a_{\rm eff} \) in parsecs, while the lower y-axis represents \( \sigma_v \) in km s\(^{-1}\).}
    }\label{fig:afit}
\end{figure}

\begin{figure}[!h]
    \centering
    \includegraphics[width=0.9\columnwidth]{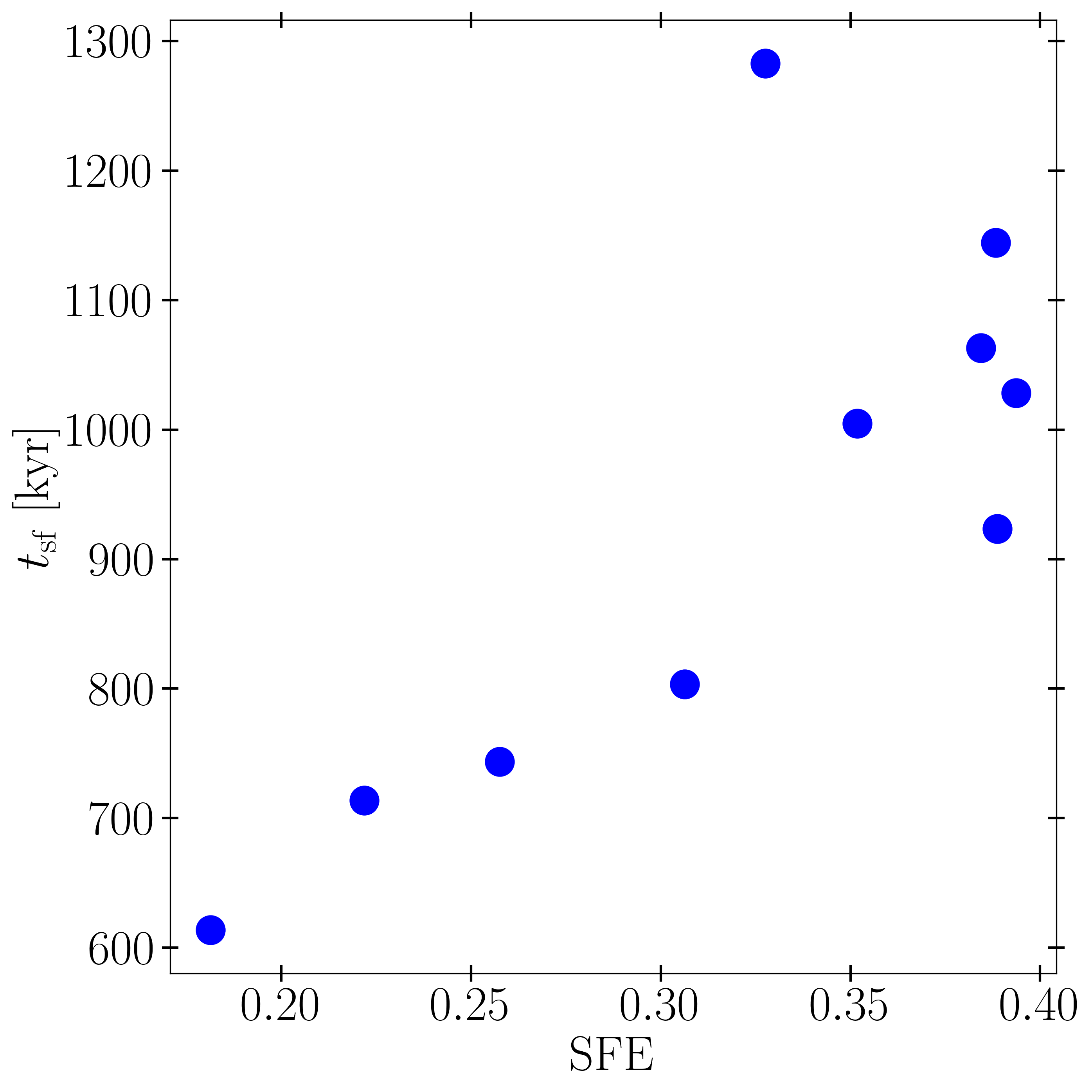}
    \caption{Dependence of the \rev{duration of} star formation \(t_{\rm sf}\) on the SFE for simulations performed at refinement level \(n=3\). Here, \(t_{\rm sf}\) is defined as the time interval from the onset of star formation until the gas mass enclosed within the radius containing half of the total stellar mass decreases to less than 20\% of the stellar mass.}
    \label{fig:half}
\end{figure}

\rev{In order to examine how stellar feedback depends on the mass of the most massive stars, we focus on the impact of the first massive star that forms with $M > 15\,M_{\odot}$, and examine how it influenced the SFE after 1~Myr. To do this, we chose four simulations with the same resolution, but different random seeds resulting in different maximum stellar masses. We ensured that within the 1 Myr interval between subsequent simulation outputs no more than one massive star was formed. (The one exception is model n3s1, where two massive stars with $M > 15\,M_{\odot}$ appeared; however, this does not contradict our analysis, as this case represents the highest stellar mass considered.) In other cases, where more than one massive star (with $M > 15\,M_{\odot}$) formed within this interval, disentangling their individual contributions would have made the analysis considerably more difficult.}

Figure~\ref{fig:SFE0_25} presents snapshots of the gas density and stellar distribution for these four simulations, which are, in order of maximum stellar mass, \texttt{n3s3}, \texttt{n3s9}, \texttt{n3s4}, and \texttt{n3s1}. The left panels correspond to the time of formation of the most massive star with $M > 15\,M_{\odot}$ in each simulation, while the right panels show a time 1~Myr later. \rev{For comparison, we highlighted the most massive stars with $M > 15\,M_{\odot}$ to include all stars likely to contribute to feedback.}

Comparison of the left and right panels in Figure~\ref{fig:SFE0_25} for each simulation shows how the formation of the most massive stars expels gas differently depending on their individual masses, the values of which are shown in the bottom right corner of each panel. Prior to the formation of the most massive stars, the gas generally collapses toward the center, forming a centrally concentrated star cluster with only limited feedback. Once the most massive stars appear, radiative and mechanical feedback begin to strongly regulate further star formation. Both the formation times of the most massive stars and the corresponding total stellar masses at these times differ (see the time and mass of the star cluster in the bottom right corner of each panel in the first column). 

We estimate the feedback during the 1~Myr time interval between the left and right panels of Figure~\ref{fig:SFE0_25} using the following approach. It is well established that stellar feedback expels gas and alters the dynamical evolution of young clusters, while stars may continue to form or be dynamically ejected \citep{2007MNRAS.380.1589B, 2009Ap&SS.324..259G}. To quantify the gas ejection, we define the feedback-induced gas mass loss \( M_{\text{loss,fb}} \) as the difference between the gas mass ejected from the grid and the mass of gas converted into stars within 5.5~pc of the cluster center. We also compute the total stellar mass formed during this period.

Figure~\ref{fig:feedback} shows the stellar mass formed and \( M_{\text{loss,fb}} \) for the \texttt{n3s3}, \texttt{n3s9}, \texttt{n3s4}, and \texttt{n3s1} simulations over the same 1~Myr time interval as the density snapshots shown in Figure~\ref{fig:SFE0_25}. 
As we can see, the stellar mass formed decreases as the mass of the most massive star increases, while \( M_{\text{loss,fb}} \) increases. These results indicate that the extremely strong feedback from the most massive stars suppresses subsequent star formation. Star formation does not cease immediately after the onset of feedback; rather, it gradually declines depending on the availability of surrounding gas and the strength of the feedback processes. 

\rev{Figure~\ref{fig:afit} shows that the maximum stellar mass \(M_{\max}\) correlates with two structural measures---the Plummer radius \(a_{\rm pl}\) and the EFF scale radius \(a_{\rm eff}\)---as well as anticorrelating with the stellar velocity dispersion \(\sigma_v\). Both the structural fits and the velocity dispersion were computed using stars within the radius around the densest region of the cluster that encloses 95\% of the total stellar mass, to provide a consistent basis of comparison. 
The point with $a_{\rm eff} = 0.0$ is excluded from the fitting, as it does not represent a physical value.
These correlations indicate that clusters hosting more massive stars are larger and dynamically colder.} Similar trends were seen in less centrally concentrated clusters by \citet{lewis2022}. These trends result primarily from the expulsion of gas in the center of the cluster prior to further star formation by feedback from the most massive star, although dynamical mass segregation can also play a role \citep{polak2025,laverde-villareal2025}.

We define a \rev{duration of} star formation \(t_{\rm sf}\) as the period from the onset of star formation until the gas mass within the half-mass radius drops below 20\% of the total stellar mass. Figure \ref{fig:half} shows the relationship between \(t_{\rm sf}\) and the SFE in simulations with \(n=3\). We see that \(t_{\rm sf}\) is positively correlated with SFE. Thus, both the \rev{duration of} formation and the mass of the most massive stars influence the final SFE of the cluster.

The influence of massive stars and their associated feedback on cluster morphology formation is consistent with the findings of \citet{lewis2022}. However, in their study, a massive star is introduced at the very beginning of cluster evolution, whereas in our simulations, massive stars emerge \rev{by chance} during the simulation.  
In contrast, for more massive molecular clouds (e.g., \(M_{\rm cloud} \sim 10^6\,M_\odot\)), stellar feedback is generally insufficient to counteract the deep gravitational potential of the cloud \citep{Polak_2024a}.

Taking into account all the results discussed above, we can answer the question we posed of what controls star formation timing and efficiency. 
\rev{As the mass of the most massive stars increases, stellar feedback becomes stronger, resulting in faster gas expulsion and a more dispersed cluster. Conversely, when the most massive stars are less massive, the gas is removed more gradually, and the resulting cluster remains more centrally concentrated. This trend is quantitatively reflected in the fitted Plummer and EFF radii, which both decrease as the mass of the most massive stars decreases. We note, however, that this conclusion applies to low-mass, low-density clouds such as those considered here; in much more massive systems, feedback from massive stars can be insufficient to halt star formation, so the regulation of efficiency proceeds differently \citep{Polak_2024a}.}

This behavior is consistent with the findings of \citet{2021MNRAS.502.6157C}, who also investigated centrally concentrated gas profiles. However, in their study, the initial gas profile is given by \( \rho(r) \propto r^{-2} \) and they used an initial gas mass of \( 8 \times 10^5\,M_{\odot} \), whereas in our simulations, the gas profile is uniform in the center and follows \( \rho(r) \propto r^{-3.3} \) beyond the Plummer radius \citep{Bek+2021}, with an initial gas mass of \( \approx 2.5 \times 10^3\,M_{\odot} \).

\subsection{Impact of resolution} 
\label{sec:Impact refinment level}

\begin{figure}[htbp]
    \centering
    \includegraphics[width=0.9\columnwidth]{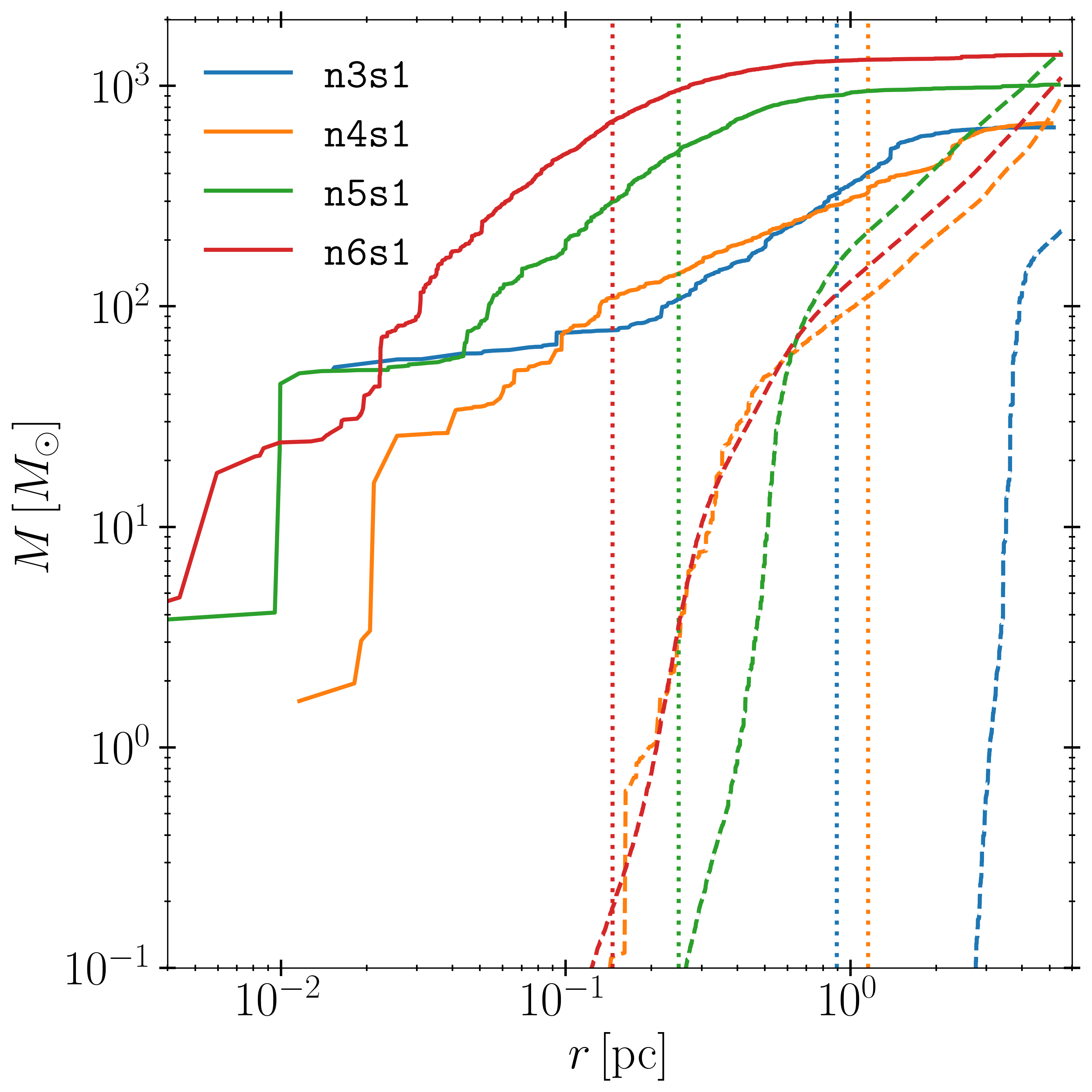}
    \caption{Enclosed mass of the star cluster ({\em solid lines}) and of the gas ({\em dashed lines}) at \(4.5\,t_{\rm ff}\) for different resolutions, with the maximum refinement level $n$ given by the first digit of the model names. The half-mass radii of the star clusters at this time are indicated for each resolution by vertical dotted lines.}
    \label{fig:cumilative}
\end{figure}

In this subsection, we examine how the numerical resolution at the finest refinement level \(n\) affects the simulation results using the same random seed by comparison of models \texttt{n3s1}, \texttt{n4s1}, \texttt{n5s1}, and \texttt{n6s1}.
Figure~\ref{fig:cumilative} 
shows the enclosed mass of stars and gas as a function of cluster radius for different refinement levels $n$. Higher resolution simulations exhibit more efficient and centrally concentrated star formation, with stars forming closer to the cluster core compared to lower-resolution runs. This outcome is driven both by the ability of higher-resolution models to resolve higher gas densities, and by the adopted star formation scheme: stars are placed within the sink radius, which decreases as resolution increases. Consequently, higher-resolution simulations naturally favor more compact stellar distributions. 

Since the rate of gas recombination is proportional to the square of the density, the ionizing luminosity from stars ionizes a smaller volume of gas at higher densities, thereby reducing the efficiency of the feedback. At the same time, the local free-fall time $t_{\rm ff}$ becomes shorter in denser regions, further accelerating central star formation. \rev{As a result, for \(n = 5\) and 6 our simulations produce compact clusters with enclosed stellar masses, shown in Figure~\ref{fig:cumilative}, that are consistent with the observations of \citet{2018MNRAS.473.4890S}, as discussed in Sect.~\ref{sec:plummer_fit}.}

\begin{figure}[htbp]
    \centering
    \includegraphics[width=0.9\columnwidth]{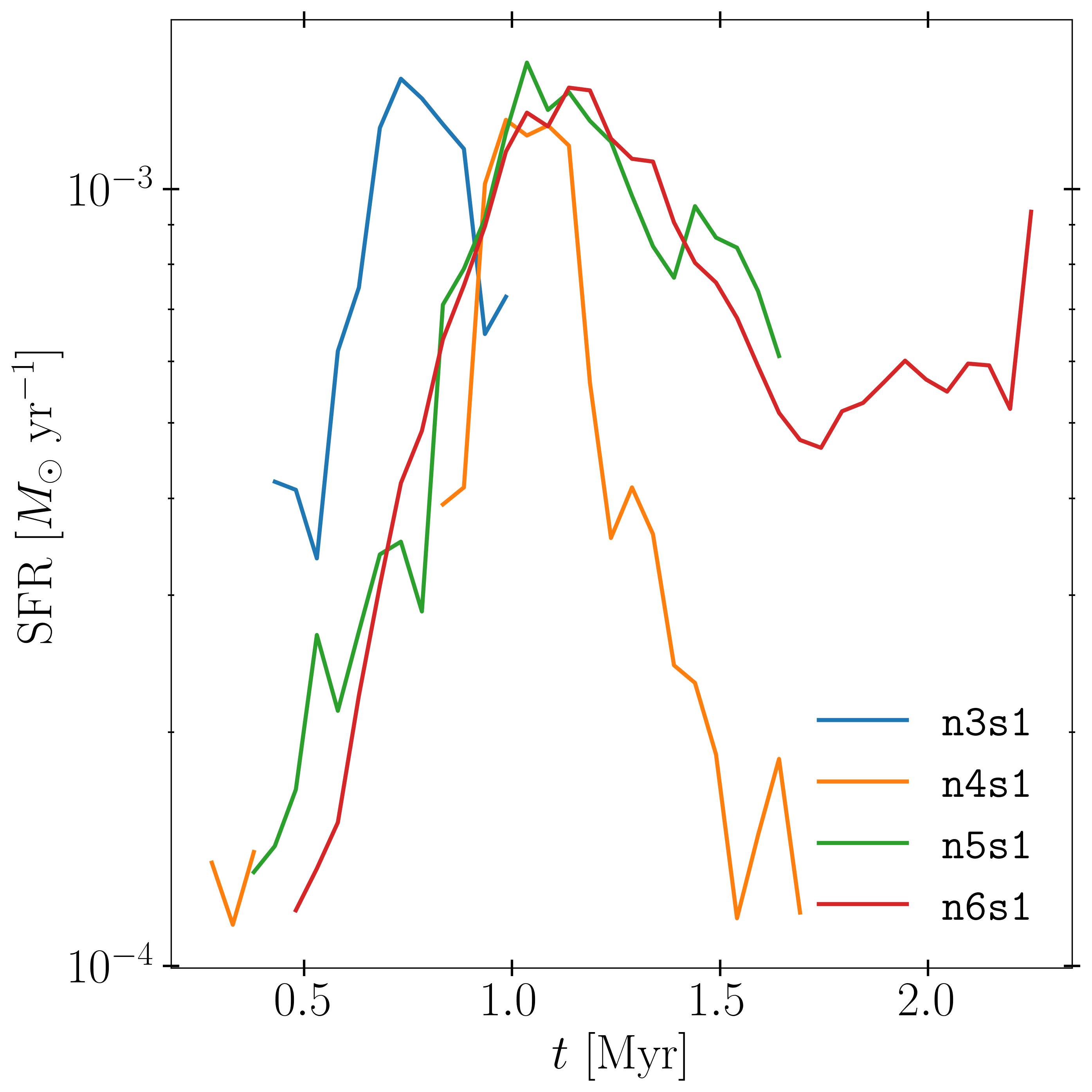}
    \caption{Star formation rate as a function of time for different resolutions, with the maximum refinement level $n$ given by the first digit of the model name. Higher resolution results in more extended and sustained star formation activity.}
    \label{fig:SFR}
\end{figure}

Figure~\ref{fig:SFR} 
shows the star formation rate (SFR) as a function of time. To emphasize differences in the star formation history, we smooth the result using a moving average with a window size of five snapshots, corresponding to $\sim 50$ kyr. For refinement level 3, star formation ceases by $\sim 1$ Myr. In contrast, star formation continues until $\sim 1.75$ Myr for refinement levels 4 and 5. At refinement level 6, a second star formation burst occurs at around 1.75 Myr, and star formation persists beyond 2 Myr.

Figure~\ref{fig:low} 
shows gas density slices centered on the densest region in the simulations as a function of resolution. This figure shows that in simulations with refinement levels 3 and 4, stellar winds from the most massive stars, with masses \(M \gtrsim 24\,M_{\odot}\), rapidly eject gas shortly after their formation, while efficient ionization by these clusters leads to the development of large ionized regions in their surroundings. In higher resolution simulations, the threshold for sink formation is larger and the accretion radius is smaller (according to the Jeans criterion). This limits accretion to the densest regions, increasing local recombination rates and increasing the cooling rate in shocked stellar winds.  As a result, stellar feedback becomes less efficient, slowing gas dispersal and producing smaller, denser, ionized regions. Refinement levels 5 and 6 can resolve \rev{filament- or sheet-like} gas structures and localized star formation (see first columns in Fig.~\ref{fig:low}). These features are absent or poorly resolved in lower-resolution runs, underscoring the importance of both accretion radius and numerical resolution for achieving physical fidelity. 

At level 6, what is the reason for the inefficiency of gas ejection and the long \rev{duration of} star formation? Simulations by \citet{2009ApJS..185..486P} report that some young clusters (even younger than 5~Myr) remain gravitationally bound to the residual gas. Several factors may explain the reduced gas expulsion at high refinement levels: the presence of turbulent gas motions \citep{2005ApJ...630..250K}, insufficient stellar feedback due to a lack of massive stars \citep{2010ApJ...710L.142F, 2010ApJ...709..191M, 2014MNRAS.442..694D}, or a deep gravitational potential from a highly concentrated gas distribution \citep{2018MNRAS.473.4890S}.

As we saw in Figure~\ref{fig:SFR} above, refinement level 3 exhibits a sharp SFR peak and earlier gas dispersal than the other refinement levels, consistent with the results of the simulation by \citet{2019MNRAS.487..364L}. They found that the central 80\% of stars tend to form near the SFR peak (see their Fig.~2), typically in the range $0.6$–$1.4\,t_{\rm ff}$, with the peak occurring when about half of the stellar mass has formed. In our case, at refinement levels other than 3, the SFR peak occurs at about 1.1 Myr, but the stellar mass does not reach half of the final cluster mass. This may be due to the strong central concentration of our initial gas, which further collapses towards the center.

In the simulations of \citet{2021MNRAS.502.6157C}, the centrally concentrated gas profile results in the SFR stopping at about 1.5 Myr (see their Fig.~3), and the peak time coincides with this same value. This may be because their MC has a large mass ($8 \times 10^5\,M_{\odot}$), the free-fall time is short, and stellar feedback is not effective due to recombination.

In summary, our results indicate that the total number and mass of stars formed seem to be approaching convergence. However, higher-resolution simulations yield more realistic ionized zones, stellar feedback, gas motion, and accretion flows.

\section{\rev{Model limitations}}
\label{sec:Model_limit}
\rev{Several limitations of our modelling approach may influence the results and their interpretation. In this section, we discuss the constraints imposed by limited resolution and computational resources, the assumption of a partially virialized initial cloud, and the omission of a strong initial magnetic field.} 

\rev{Simulations with low refinement levels restrict the ability to resolve fine-scale gas structures, such as filaments or sheets, which are better captured at higher refinement levels. To mitigate this limitation, we also performed complementary simulations at higher resolutions. At lower refinement levels, larger sink accretion radii (\(r_{\rm sink} = 0.64\,\mathrm{pc}\) for \(n=3\)) allow star formation to occur in less dense regions, artificially enhancing feedback efficiency. By contrast, higher-resolution simulations with smaller sink radii (\(r_{\rm sink} = 0.08\,\mathrm{pc}\) for \(n=6\)) resolve denser gas more accurately and produce more realistic ionized zones (see Section~\ref{sec:Impact refinment level} for details).}

\rev{The high computational cost of simulations at refinement levels \(n=5\) and \(n=6\) limited the number of realizations we could perform, particularly with different random seeds. At \(n=3\), we conducted ten realizations (\texttt{n3s1--10}) to capture statistical variations from stochastic IMF sampling, whereas for \(n=4\)--6, only a single realization was feasible due to computational constraints (Table~\ref{tab:deluxesplit}). For example, simulations at refinement levels \(n=3\), 4, 5, and 6 required approximately 7 days, 45 days, 4 months, and 7 months, respectively. At \(t=4.5\,t_{\rm ff}\), the total number of grid cells reached \(3\times10^5\), \(2\times10^6\), \(9\times10^6\), and \(3\times10^7\) cells, respectively. Moreover, for \(n=6\), the simulations could not be run on fewer than 140 computing cores, and toward the end the code advanced only \(\sim 10\,\mathrm{kyr}\) every three days.}

\begin{figure*}[htbp]
    \centering
    \includegraphics[width=0.9\textwidth]{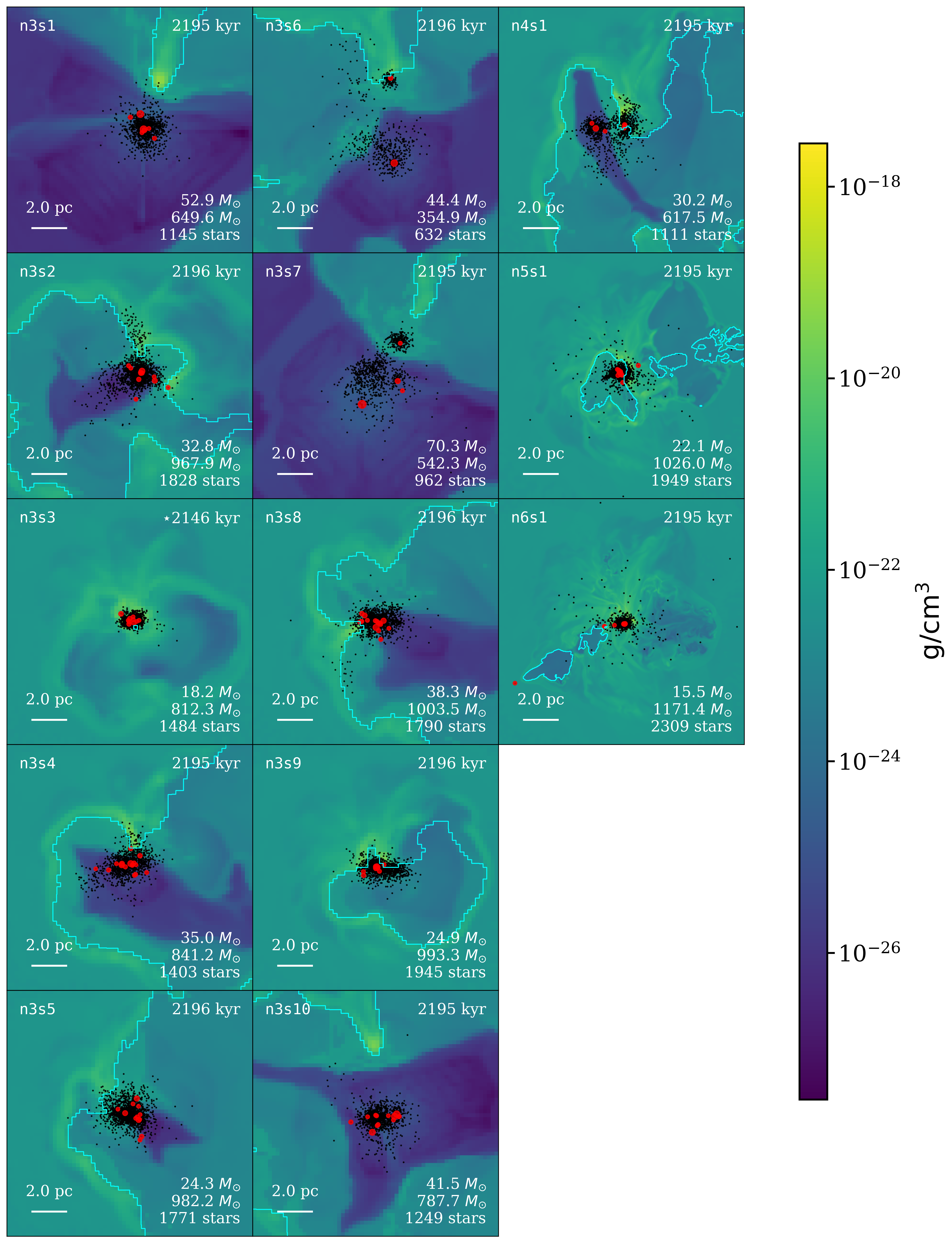}
    \caption{\rev{Snapshots of gas density and stellar distribution at \(4\,t_{\rm ff}\) for different maximum grid resolutions. The ionization fronts are marked with cyan lines. Stars are shown as black dots; massive stars are highlighted in red, with sizes proportional to their masses. In each panel, the following notations are used: {\em (bottom left)} physical scale; {\em (bottom right)} the mass of the most massive star, total stellar mass, and number of stars; {\em (top left)} model label; {\em (top right)} simulation time. The color scale indicates the gas density in g~cm\(^{-3}\). The panel for model \texttt{n3s3}, marked with a $\star$ (simulation time), is shown 40~kyr earlier than \(4\,t_{\rm ff}\), because a \(60\,M_\odot\) star forms shortly afterward; at the displayed time, the feedback is dominated by an \(18\,M_\odot\) star.}}
    \label{fig:all_dens}
\end{figure*}

\rev{Figure~\ref{fig:all_dens} presents gas density slices centered on the densest regions as a function of resolution. The figure shows that the overall gas dynamics are governed primarily by stellar feedback. Across different refinement levels, the evolution of gas and the onset of star formation proceed in a broadly similar manner, depending mainly on the mass of the most massive stars (compare the ionized regions in models \texttt{n3s2} versus \texttt{n4s1}, \texttt{n3s9} versus \texttt{n5s1}, and \texttt{n3s3} versus \texttt{n6s1}). This suggests that the qualitative results obtained at refinement level \(n=3\) can be considered robust. Furthermore, sub-cluster formation occurs not only at \(n=3\) but also at higher resolutions, such as \(n=4\) (model \texttt{n4s1}). In model \texttt{n3s3}, the snapshot indicated by a $\star$ is shown 40~kyr earlier than \(4\,t_{\rm ff}\), since a very massive star of \(60\,M_\odot\) forms shortly afterward. At the displayed epoch, however, the primary feedback effect arises from an \(18\,M_\odot\) star, which governs the gas evolution at that stage.}

\rev{The deviation of our stellar cluster profiles from analytical models may partly result from only balancing kinetic and gravitational energy globally but not locally (see Sect.~\ref{InitialConditions})
and from the neglect of strong initial magnetic fields \citep{2011A&A...528A..72H,2012ApJ...761..156F,2016A&A...593L..14F,2020ApJ...891..168W}. The initial cloud is set up with a virial parameter \(\alpha_{\rm vir} = 0.5\), indicating that the total turbulent kinetic energy is half the gravitational potential energy (Section~\ref{InitialConditions}). This does not fully capture the energy balance of the initialized clouds, though, since centrally concentrated clouds require stronger central turbulent support to balance their deep gravitational potentials.
}

\rev{A fully virialized cloud could delay global collapse, extend the \rev{duration of} star formation (\(t_{\rm sf}\)), and increase SFE by providing greater support against rapid central collapse. This would likely produce more spatially distributed star formation and less compact stellar clusters. However, unless there were a physical mechanism to drive the turbulence in a centrally concentrated matter, the turbulence would decay in a free-fall time \citep{mac-low1999a} and the cloud would then evolve without much change.}

\rev{Indeed, numerical studies have shown that clouds with higher virial parameters undergo more fragmented star formation, leading to the formation of multiple sub-clusters rather than a single, centrally concentrated cluster \citep{2003MNRAS.343..413B,2005MNRAS.359..809C,2012ApJ...761..156F}. In our case, increasing \(\alpha_{\rm vir}\) could promote sub-cluster formation (as shown in \texttt{n3s6}, \texttt{n3s7}, and \texttt{n4s1} by Figure~\ref{fig:Fbound}), yielding stellar density profiles that deviate from the analytical Plummer model (\(a_\star = 1.1\,\mathrm{pc}\)) and resemble EFF profiles with variable \(\gamma_{\rm eff}\) (Section~\ref{sec:plummer_fit}). Such conditions would likely increase the fitted scale radii (\(a_{\rm pl}\), \(a_{\rm eff}\)), producing less compact clusters.}

\rev{Our simulations neglect the role of strong magnetic fields, which is a significant simplification. Magnetic fields provide additional support against gravitational collapse, influence gas dynamics, and modulate the efficiency of stellar feedback. Previous studies have shown that magnetic fields can reduce the fragmentation of molecular clouds, delay star formation, and alter the morphology of ionized regions \citep{2007ARA&A..45..565M,2011ApJ...730...40P,2015MNRAS.450.4035F}. Including stronger magnetic fields in our models could therefore lead to reduced star formation rates, larger characteristic scales of stellar clustering, and modified feedback-driven gas expulsion.}

\rev{In summary, our results are shaped by several key modeling assumptions and constraints. Limited resolution and high computational cost restrict the number of realizations at the highest refinement levels, though comparisons across resolutions suggest that low-resolution runs capture the essential physics of feedback-driven gas dispersal. The assumption of a partially virialized initial cloud likely biases our clusters toward higher central concentrations than might occur in fully virialized systems, where turbulence would delay collapse and promote greater sub-clustering. Finally, the omission of strong magnetic fields removes an important source of support and regulation of fragmentation, potentially leading to higher star formation rates and more compact stellar structures than would otherwise be expected.}

\section{Conclusion}
\label{sec:Conclusion}

In this study, we investigated how massive star formation and gas structure influence the formation and evolution of young star clusters. Using simulations with varying numerical resolution and massive star formation occurrence, we followed the collapse, star formation, and feedback processes to determine their role in regulating cluster properties. 

\rev{The evolution of our simulations can be summarized as follows: gravitational collapse begins \rev{globally towards the center of the cloud}, \rev{possibly} leading to the formation of \rev{sub-structures resembling filaments or sheets} and the first stars \rev{appear} at approximately \(1\,t_{\rm ff}\); accretion then continues in the central region, fuelling further star formation; this process becomes progressively less efficient as the most massive stars form, marking the onset of stellar feedback.} 
We find the following conclusions.

\begin{itemize}
    \item The gas in the central regions has shorter free-fall times, leading to rapid collapse and intense star formation, while gas in the outer regions has longer free-fall times (see Fig.~\ref{fig:prfle}), so it cannot form stars prior to falling in to the center.
    \item \rev{As a result, star formation produces rather more centrally concentrated profiles (Fig.~\ref{fig:Prfle_b}) predicted from the initial gas distribution by the local collapse theory of \citet{2013A&A...549A.132P}.}  
    \item Stars \rev{often form in sub-clusters along filament- or sheet-like gas structures even in the highly centralized initial gas distributions studied here. Early-forming massive stars promote the formation of interacting sub-clusters (see Fig.~\ref{fig:Fbound}). In some models, the sub-clusters merge into a single cluster, while in others they survive separately.  The evolution of such sub-clusters in less centrally concentrated gas was described by \citet{cournoyer-cloutier2023}.}
    \item \rev{The 
    masses of the most massive stars are critical in determining the strength of stellar feedback and shaping the evolution of young star clusters (Fig.~\ref{fig:SFE0_25}--\ref{fig:afit}). This extends the results from \citet{lewis2022} to more centrally concentrated initial conditions and a range of massive star formation times.}
    \item \rev{The final SFE is influenced by both the mass of the most massive star formed and the \rev{duration of} star formation, $t_{\rm sf}$ (see Figs.~\ref{fig:SFE0_25} and~\ref{fig:half}).}
    \item \rev{Our convergence study shows that the total number and mass of stars formed appears to be approaching convergence at our highest grid resolutions (Figs.~\ref{fig:cumilative} and~\ref{fig:SFR}). However, higher grid resolution continues to improve the modeling of ionized zones, stellar feedback, and accretion in dense regions (Fig.~\ref{fig:all_dens}).} 
\end{itemize}
\rev{However, these conclusions should be interpreted in light of the limitations of our models, which include assumptions about numerical resolution, cloud virialization, and the absence of strong initial magnetic fields. A detailed discussion of these caveats is provided in Section~\ref{sec:Model_limit}. Addressing these limitations in future work will be crucial for building a more comprehensive picture of star cluster formation in realistic molecular cloud environments.}

\section*{\rev{Data availability}}

\rev{All density visualization videos and corresponding snapshot files generated from the simulations and used in this study 
are publicly available at \href{https://doi.org/10.5281/zenodo.17005409}{Zenodo}\footnote{\url{https://doi.org/10.5281/zenodo.17005409}}.}

\begin{acknowledgements}
\rev{We thank the anonymous referee for the constructive comments and suggestions that helped improve the clarity and quality of this paper. The authors thank S. Portegies Zwart, S. L. McMillan, E. P. Andersson, and S. M. Appel for scientific discussions and extensive support in our use of the AMUSE and Torch software.} This research has been funded by the Science Committee of the Ministry of Science and Higher Education, Republic of Kazakhstan (Grant No. AP19677351).
AA acknowledges support from a Bolashaq International Scholarship. M-MML and BP acknowledge support from US
NSF grant AST23-07950. CC-C is supported by a Canada Graduate Scholarship--Doctoral (CGS D) from NSERC. EA's work is supported by Nazarbayev University Faculty Development Competitive Research Grant Program (no. 040225FD4713). Some computations presented here were performed on Snellius through the Dutch National Supercomputing Center SURF grants 15220 and 2023/ENW/01498863. M-MML thanks the Institut f{\"u}r Theoretische Astrophysik of Heidelberg University for hospitality during work on this paper. 
 
\end{acknowledgements}

\bibliography{sample631,maclow}

\begin{thebibliography}{131}
\expandafter\ifx\csname natexlab\endcsname\relax\def\natexlab#1{#1}\fi

\bibitem[{{Adams}(2000)}]{2000ApJ...542..964A}
{Adams}, F.~C. 2000, \apj, 542, 964

\bibitem[{{Agertz} {et~al.}(2013){Agertz}, {Kravtsov}, {Leitner}, \&
  {Gnedin}}]{2013ApJ...770...25A}
{Agertz}, O., {Kravtsov}, A.~V., {Leitner}, S.~N., \& {Gnedin}, N.~Y. 2013,
  \apj, 770, 25

\bibitem[{{Alina} {et~al.}(2019){Alina}, {Ristorcelli}, {Montier},
  {Abdikamalov}, {Juvela}, {Ferri{\`e}re}, {Bernard}, \&
  {Micelotta}}]{Alina+2019}
{Alina}, D., {Ristorcelli}, I., {Montier}, L., {et~al.} 2019, \mnras, 485, 2825

\bibitem[{{Baczynski} {et~al.}(2015){Baczynski}, {Glover}, \&
  {Klessen}}]{baczynski2015}
{Baczynski}, C., {Glover}, S.~C.~O., \& {Klessen}, R.~S. 2015, \mnras, 454, 380

\bibitem[{{Bate} {et~al.}(1995){Bate}, {Bonnell}, \& {Price}}]{bate1995}
{Bate}, M.~R., {Bonnell}, I.~A., \& {Price}, N.~M. 1995, \mnras, 277, 362

\bibitem[{{Baumgardt} \& {Kroupa}(2007)}]{2007MNRAS.380.1589B}
{Baumgardt}, H. \& {Kroupa}, P. 2007, \mnras, 380, 1589

\bibitem[{{Bissekenov} {et~al.}(2024{\natexlab{a}}){Bissekenov}, {Kalambay},
  {Abdikamalov}, {Pang}, {Berczik}, \& {Shukirgaliyev}}]{Abylay+2024}
{Bissekenov}, A., {Kalambay}, M., {Abdikamalov}, E., {et~al.}
  2024{\natexlab{a}}, \aap, 689, A282

\bibitem[{{Bissekenov} {et~al.}(2024{\natexlab{b}}){Bissekenov}, Kalambay,
  Abylkairov, \& Shukirgaliyev}]{bissekenov2024exploring}
{Bissekenov}, A., Kalambay, M.~T., Abylkairov, Y.~S., \& Shukirgaliyev, B.~T.
  2024{\natexlab{b}}, Recent Contributions to Physics, 90

\bibitem[{{Bissekenov} {et~al.}(2025){Bissekenov}, {Pang}, {Kamlah},
  {Kouwenhoven}, {Spurzem}, {Shukirgaliyev}, {Giersz}, {Askar}, \&
  {Berczik}}]{Abylay+2025}
{Bissekenov}, A., {Pang}, X., {Kamlah}, A., {et~al.} 2025, A\&A, 699, A196

\bibitem[{{Bonnell} {et~al.}(2003){Bonnell}, {Bate}, \&
  {Vine}}]{2003MNRAS.343..413B}
{Bonnell}, I.~A., {Bate}, M.~R., \& {Vine}, S.~G. 2003, \mnras, 343, 413

\bibitem[{Brinkmann {et~al.}(2017)Brinkmann, Banerjee, Motwani, \&
  Kroupa}]{Brinkmann2016}
Brinkmann, N., Banerjee, S., Motwani, B., \& Kroupa, P. 2017, Astronomy {\&}
  Astrophysics, 600, A49

\bibitem[{{Burki}(1978)}]{1978A&A....62..159B}
{Burki}, G. 1978, \aap, 62, 159

\bibitem[{{Canning} {et~al.}(2014){Canning}, {Ryon}, {Gallagher}, {Kotulla},
  {O'Connell}, {Fabian}, {Johnstone}, {Conselice}, {Hicks}, {Rosario}, \&
  {Wyse}}]{2014MNRAS.444..336C}
{Canning}, R.~E.~A., {Ryon}, J.~E., {Gallagher}, J.~S., {et~al.} 2014, \mnras,
  444, 336

\bibitem[{{Chen} \& {Ko}(2009)}]{2009ApJ...698.1659C}
{Chen}, H.-C. \& {Ko}, C.-M. 2009, \apj, 698, 1659

\bibitem[{{Chen} {et~al.}(2021){Chen}, {Li}, \&
  {Vogelsberger}}]{2021MNRAS.502.6157C}
{Chen}, Y., {Li}, H., \& {Vogelsberger}, M. 2021, \mnras, 502, 6157

\bibitem[{{Clark} {et~al.}(2005){Clark}, {Bonnell}, {Zinnecker}, \&
  {Bate}}]{2005MNRAS.359..809C}
{Clark}, P.~C., {Bonnell}, I.~A., {Zinnecker}, H., \& {Bate}, M.~R. 2005,
  \mnras, 359, 809

\bibitem[{{Colella} \& {Woodward}(1984)}]{1984JCoPh..54..174C}
{Colella}, P. \& {Woodward}, P.~R. 1984, Journal of Computational Physics, 54,
  174

\bibitem[{{Cournoyer-Cloutier} {et~al.}(2023){Cournoyer-Cloutier}, {Sills},
  {Harris}, {Appel}, {Lewis}, {Polak}, {Tran}, {Wilhelm}, {Mac Low},
  {McMillan}, \& {Portegies Zwart}}]{cournoyer-cloutier2023}
{Cournoyer-Cloutier}, C., {Sills}, A., {Harris}, W.~E., {et~al.} 2023, \mnras,
  521, 1338

\bibitem[{{Cournoyer-Cloutier} {et~al.}(2024){Cournoyer-Cloutier}, {Sills},
  {Harris}, {Polak}, {Rieder}, {Andersson}, {Appel}, {Mac Low}, {McMillan}, \&
  {Portegies Zwart}}]{cournoyer-cloutier2024}
{Cournoyer-Cloutier}, C., {Sills}, A., {Harris}, W.~E., {et~al.} 2024, ApJ,
  977, 203

\bibitem[{{Cournoyer-Cloutier} {et~al.}(2021){Cournoyer-Cloutier}, {Tran},
  {Lewis}, {Wall}, {Harris}, {Mac Low}, {McMillan}, {Portegies Zwart}, \&
  {Sills}}]{cournoyer-cloutier2021}
{Cournoyer-Cloutier}, C., {Tran}, A., {Lewis}, S., {et~al.} 2021, MNRAS, 501,
  4464

\bibitem[{{Dale} {et~al.}(2025){Dale}, {Graham}, {Barnes}, {Baron}, {Bigiel},
  {Boquien}, {Chandar}, {Chastenet}, {Chown}, {Egorov}, {Glover}, {Hands},
  {Henny}, {Indebetouw}, {Klessen}, {Larson}, {Lee}, {Leroy}, {Maschmann},
  {Pathak}, {Rodr{\'\i}guez}, {Rosolowsky}, {Sandstrom}, {Schinnerer},
  {Sutter}, {Thilker}, {Weinbeck}, {Whitmore}, {Williams}, \&
  {Wofford}}]{2025AJ....169..133D}
{Dale}, D.~A., {Graham}, G.~B., {Barnes}, A.~T., {et~al.} 2025, \aj, 169, 133

\bibitem[{{Dale} {et~al.}(2014){Dale}, {Ngoumou}, {Ercolano}, \&
  {Bonnell}}]{2014MNRAS.442..694D}
{Dale}, J.~E., {Ngoumou}, J., {Ercolano}, B., \& {Bonnell}, I.~A. 2014, \mnras,
  442, 694

\bibitem[{{Draine}(2011)}]{draine2011}
{Draine}, B.~T. 2011, {Physics of the Interstellar and Intergalactic Medium}
  (Princeton: Princeton U. Press)

\bibitem[{Dubey {et~al.}(2014)Dubey, Antypas, Calder, Daley, Fryxell,
  Gallagher, Lamb, Lee, Olson, Reid, Rich, Ricker, Riley, Rosner, Siegel,
  Taylor, Weide, Timmes, Vladimirova, \& ZuHone}]{dubey2014}
Dubey, A., Antypas, K., Calder, A.~C., {et~al.} 2014, The International Journal
  of High Performance Computing Applications, 28, 225

\bibitem[{{Elmegreen} \& {Scalo}(2004)}]{2004ARA&A..42..211E}
{Elmegreen}, B.~G. \& {Scalo}, J. 2004, \araa, 42, 211

\bibitem[{{Elson} {et~al.}(1987){Elson}, {Fall}, \& {Freeman}}]{Elson1987}
{Elson}, R. A.~W., {Fall}, S.~M., \& {Freeman}, K.~C. 1987, \apj, 323, 54

\bibitem[{{Fall} {et~al.}(2010){Fall}, {Krumholz}, \&
  {Matzner}}]{2010ApJ...710L.142F}
{Fall}, S.~M., {Krumholz}, M.~R., \& {Matzner}, C.~D. 2010, \apjl, 710, L142

\bibitem[{{Farias} \& {Tan}(2023)}]{2023MNRAS.523.2083F}
{Farias}, J.~P. \& {Tan}, J.~C. 2023, \mnras, 523, 2083

\bibitem[{{Farias} {et~al.}(2019){Farias}, {Tan}, \&
  {Chatterjee}}]{2019MNRAS.483.4999F}
{Farias}, J.~P., {Tan}, J.~C., \& {Chatterjee}, S. 2019, \mnras, 483, 4999

\bibitem[{{Federrath}(2015)}]{2015MNRAS.450.4035F}
{Federrath}, C. 2015, \mnras, 450, 4035

\bibitem[{{Federrath} {et~al.}(2010){Federrath}, {Banerjee}, {Clark}, \&
  {Klessen}}]{federrath2010}
{Federrath}, C., {Banerjee}, R., {Clark}, P.~C., \& {Klessen}, R.~S. 2010,
  \apj, 713, 269

\bibitem[{{Federrath} \& {Klessen}(2012)}]{2012ApJ...761..156F}
{Federrath}, C. \& {Klessen}, R.~S. 2012, \apj, 761, 156

\bibitem[{{Feigelson} {et~al.}(2005){Feigelson}, {Getman}, {Townsley},
  {Garmire}, {Preibisch}, {Grosso}, {Montmerle}, {Muench}, \&
  {McCaughrean}}]{2005ApJS..160..379F}
{Feigelson}, E.~D., {Getman}, K., {Townsley}, L., {et~al.} 2005, \apjs, 160,
  379

\bibitem[{{Fierlinger} {et~al.}(2016){Fierlinger}, {Burkert}, {Ntormousi},
  {Fierlinger}, {Schartmann}, {Ballone}, {Krause}, \&
  {Diehl}}]{Fierlinger+2016}
{Fierlinger}, K.~M., {Burkert}, A., {Ntormousi}, E., {et~al.} 2016, \mnras,
  456, 710

\bibitem[{{Fontani} {et~al.}(2016){Fontani}, {Commer{\c{c}}on}, {Giannetti},
  {Beltr{\'a}n}, {S{\'a}nchez-Monge}, {Testi}, {Brand}, {Caselli}, {Cesaroni},
  {Dodson}, {Longmore}, {Rioja}, {Tan}, \& {Walmsley}}]{2016A&A...593L..14F}
{Fontani}, F., {Commer{\c{c}}on}, B., {Giannetti}, A., {et~al.} 2016, \aap,
  593, L14

\bibitem[{{Fryxell} {et~al.}(2000){Fryxell}, {Olson}, {Ricker}, {Timmes},
  {Zingale}, {Lamb}, {MacNeice}, {Rosner}, {Truran}, \& {Tufo}}]{fryxell2000}
{Fryxell}, B., {Olson}, K., {Ricker}, P., {et~al.} 2000, \apjs, 131, 273

\bibitem[{{Fujii} {et~al.}(2007){Fujii}, {Iwasawa}, {Funato}, \&
  {Makino}}]{fujii2007}
{Fujii}, M., {Iwasawa}, M., {Funato}, Y., \& {Makino}, J. 2007, \pasj, 59, 1095

\bibitem[{{Goodwin}(1997)}]{1997MNRAS.284..785G}
{Goodwin}, S.~P. 1997, \mnras, 284, 785

\bibitem[{{Goodwin}(2009)}]{2009Ap&SS.324..259G}
{Goodwin}, S.~P. 2009, \apss, 324, 259

\bibitem[{{Goodwin} \& {Bastian}(2006)}]{2006MNRAS.373..752G}
{Goodwin}, S.~P. \& {Bastian}, N. 2006, \mnras, 373, 752

\bibitem[{{Grudi{\'c}} {et~al.}(2018){Grudi{\'c}}, {Hopkins},
  {Faucher-Gigu{\`e}re}, {Quataert}, {Murray}, \&
  {Kere{\v{s}}}}]{2018MNRAS.475.3511G}
{Grudi{\'c}}, M.~Y., {Hopkins}, P.~F., {Faucher-Gigu{\`e}re}, C.-A., {et~al.}
  2018, \mnras, 475, 3511

\bibitem[{{Gusev} {et~al.}(2016){Gusev}, {Sakhibov}, {Piskunov}, {Kharchenko},
  {Pilyugin}, {Ezhkova}, {Khramtsova}, {Guslyakova}, {Bruevich}, {Dodonov},
  {Lang}, {Shimanovskaya}, \& {Efremov}}]{2016A&AT...29..293G}
{Gusev}, A.~S., {Sakhibov}, F.~K., {Piskunov}, A.~E., {et~al.} 2016,
  Astronomical and Astrophysical Transactions, 29, 293

\bibitem[{{H{\"a}rer} {et~al.}(2025){H{\"a}rer}, {Vieu}, \&
  {Reville}}]{2025arXiv250319745H}
{H{\"a}rer}, L., {Vieu}, T., \& {Reville}, B. 2025, arXiv e-prints,
  arXiv:2503.19745

\bibitem[{{Haworth} {et~al.}(2018){Haworth}, {Glover}, {Koepferl}, {Bisbas}, \&
  {Dale}}]{2018NewAR..82....1H}
{Haworth}, T.~J., {Glover}, S. C.~O., {Koepferl}, C.~M., {Bisbas}, T.~G., \&
  {Dale}, J.~E. 2018, \nar, 82, 1

\bibitem[{{Heitsch} {et~al.}(2001){Heitsch}, {Mac Low}, \&
  {Klessen}}]{2001ApJ...547..280H}
{Heitsch}, F., {Mac Low}, M.-M., \& {Klessen}, R.~S. 2001, \apj, 547, 280

\bibitem[{{Hennebelle} {et~al.}(2011){Hennebelle}, {Commer{\c{c}}on}, {Joos},
  {Klessen}, {Krumholz}, {Tan}, \& {Teyssier}}]{2011A&A...528A..72H}
{Hennebelle}, P., {Commer{\c{c}}on}, B., {Joos}, M., {et~al.} 2011, \aap, 528,
  A72

\bibitem[{{Heydari-Malayeri} {et~al.}(2003){Heydari-Malayeri}, {Charmandaris},
  {Deharveng}, {Meynadier}, {Rosa}, {Schaerer}, \&
  {Zinnecker}}]{2003IAUS..212..553H}
{Heydari-Malayeri}, M., {Charmandaris}, V., {Deharveng}, L., {et~al.} 2003, in
  IAU Symposium, Vol. 212, A Massive Star Odyssey: From Main Sequence to
  Supernova, ed. K.~{van der Hucht}, A.~{Herrero}, \& C.~{Esteban} (San
  Francisco: Astron.\ Soc.\ Pacific), 553

\bibitem[{{Hopkins} {et~al.}(2012){Hopkins}, {Quataert}, \&
  {Murray}}]{Hopkins+2012}
{Hopkins}, P.~F., {Quataert}, E., \& {Murray}, N. 2012, \mnras, 421, 3522

\bibitem[{{Hut} {et~al.}(1995){Hut}, {Makino}, \& {McMillan}}]{Hut1995}
{Hut}, P., {Makino}, J., \& {McMillan}, S. 1995, \apjl, 443, L93

\bibitem[{{Ib{\'a}{\~n}ez-Mej{\'\i}a}
  {et~al.}(2016){Ib{\'a}{\~n}ez-Mej{\'\i}a}, {Mac Low}, {Klessen}, \&
  {Baczynski}}]{2016ApJ...824...41I}
{Ib{\'a}{\~n}ez-Mej{\'\i}a}, J.~C., {Mac Low}, M.-M., {Klessen}, R.~S., \&
  {Baczynski}, C. 2016, \apj, 824, 41

\bibitem[{{Ib{\'a}{\~n}ez-Mej{\'\i}a}
  {et~al.}(2017){Ib{\'a}{\~n}ez-Mej{\'\i}a}, {Mac Low}, {Klessen}, \&
  {Baczynski}}]{2017ApJ...850...62I}
{Ib{\'a}{\~n}ez-Mej{\'\i}a}, J.~C., {Mac Low}, M.-M., {Klessen}, R.~S., \&
  {Baczynski}, C. 2017, \apj, 850, 62

\bibitem[{{Ishchenko} {et~al.}(2025){Ishchenko}, {Masliukh}, {Hradov},
  {Berczik}, {Shukirgaliyev}, \& {Omarov}}]{Marina+2025}
{Ishchenko}, M., {Masliukh}, V., {Hradov}, M., {et~al.} 2025, \aap, 694, A33

\bibitem[{{Jeans}(1902)}]{jeans1902}
{Jeans}, J.~H. 1902, Roy.\ Soc.\ Lond.\ Phil.\ Trans.\ Ser.\ A, 199, 1

\bibitem[{Kalambay {et~al.}(2022)Kalambay, Naurzbayeva, Otebay, Abdinassilimm,
  Kuvatova, Assilkhan, Panamarev, Shukirgaliyev, \& Berczik}]{Kalambay2022}
Kalambay, M.~T., Naurzbayeva, A.~Z., Otebay, A.~B., {et~al.} 2022, Recent
  Contributions to Physics, 83, 4

\bibitem[{Kalambay {et~al.}(2025)Kalambay, Otebay, Nazar, Assilkhan, \&
  Shukirgaliyev}]{Kalambay2025}
Kalambay, M.~T., Otebay, A.~B., Nazar, A.~B., Assilkhan, A., \& Shukirgaliyev,
  B.~T. 2025, Herald of the Kazakh-British Technical University, 22, 312

\bibitem[{{Karam} \& {Sills}(2022)}]{Jeremy2022}
{Karam}, J. \& {Sills}, A. 2022, \mnras, 513, 6095

\bibitem[{{Karam} \& {Sills}(2024)}]{2024ApJ...967...86K}
{Karam}, J. \& {Sills}, A. 2024, \apj, 967, 86

\bibitem[{{Kim} {et~al.}(2018){Kim}, {Kim}, \& {Ostriker}}]{Kim+2018}
{Kim}, J.-G., {Kim}, W.-T., \& {Ostriker}, E.~C. 2018, \apj, 859, 68

\bibitem[{{Kolmogorov}(1941)}]{1941DoSSR..30..301K}
{Kolmogorov}, A. 1941, Akademiia Nauk SSSR Doklady, 30, 301

\bibitem[{{Komesh} {et~al.}(2024){Komesh}, {Garay}, {Henkel}, {Omar},
  {Estalella}, {Assembay}, {Li}, {Guzm{\'a}n}, {Esimbek}, {Huang}, {He},
  {Alimgazinova}, {Kyzgarina}, {Bekdaulet}, {Zhumabay}, \&
  {Manapbayeva}}]{Komesh+2024}
{Komesh}, T., {Garay}, G., {Henkel}, C., {et~al.} 2024, \apj, 967, 15

\bibitem[{{Kroupa}(2002)}]{2002Sci...295...82K}
{Kroupa}, P. 2002, Science, 295, 82

\bibitem[{{Krumholz} \& {McKee}(2005)}]{2005ApJ...630..250K}
{Krumholz}, M.~R. \& {McKee}, C.~F. 2005, \apj, 630, 250

\bibitem[{{Krumholz} {et~al.}(2004){Krumholz}, {McKee}, \&
  {Klein}}]{krumholz2004}
{Krumholz}, M.~R., {McKee}, C.~F., \& {Klein}, R.~I. 2004, \apj, 611, 399

\bibitem[{{Kuhn} {et~al.}(2015){Kuhn}, {Feigelson}, {Getman}, {Sills}, {Bate},
  \& {Borissova}}]{2015ApJ...812..131K}
{Kuhn}, M.~A., {Feigelson}, E.~D., {Getman}, K.~V., {et~al.} 2015, \apj, 812,
  131

\bibitem[{{Lada} \& {Lada}(2003)}]{2003ARA&A..41...57L}
{Lada}, C.~J. \& {Lada}, E.~A. 2003, \araa, 41, 57

\bibitem[{{Lada} {et~al.}(1984){Lada}, {Margulis}, \&
  {Dearborn}}]{1984ApJ...285..141L}
{Lada}, C.~J., {Margulis}, M., \& {Dearborn}, D. 1984, \apj, 285, 141

\bibitem[{{Lanz} \& {Hubeny}(2003)}]{lanz2003}
{Lanz}, T. \& {Hubeny}, I. 2003, \apjs, 146, 417

\bibitem[{{Larson}(1981)}]{1981MNRAS.194..809L}
{Larson}, R.~B. 1981, \mnras, 194, 809

\bibitem[{{Laverde-Villarreal} {et~al.}(2025){Laverde-Villarreal}, {Sills},
  {Cournoyer-Cloutier}, \& {Arias Callejas}}]{laverde-villareal2025}
{Laverde-Villarreal}, E., {Sills}, A., {Cournoyer-Cloutier}, C., \& {Arias
  Callejas}, V. 2025, accepted to \apj, arXiv:2507.00815

\bibitem[{{Leisawitz} {et~al.}(1989){Leisawitz}, {Bash}, \&
  {Thaddeus}}]{1989ApJS...70..731L}
{Leisawitz}, D., {Bash}, F.~N., \& {Thaddeus}, P. 1989, \apjs, 70, 731

\bibitem[{Levenberg(1944)}]{Levenberg1944}
Levenberg, K. 1944, Quarterly of Applied Mathematics, 2, 164

\bibitem[{Lewis {et~al.}(2022)Lewis, {McMillan}, {Mac Low},
  {Cournoyer-Cloutier}, Polak, Wilhelm, Tran, Sills, {Portegies Zwart},
  {Klessen}, \& Wall}]{lewis2022}
Lewis, S.~C., {McMillan}, S.~L.~W., {Mac Low}, M.~M., {et~al.} 2022, \apj,
  subm.

\bibitem[{{Lewis} {et~al.}(2023){Lewis}, {McMillan}, {Mac Low},
  {Cournoyer-Cloutier}, {Polak}, {Wilhelm}, {Tran}, {Sills}, {Portegies Zwart},
  {Klessen}, \& {Wall}}]{2023AAS...24110903L}
{Lewis}, S.~C., {McMillan}, S. L.~W., {Mac Low}, M.-M., {et~al.} 2023, \apj,
  944, 211

\bibitem[{{Li} {et~al.}(2019){Li}, {Vogelsberger}, {Marinacci}, \&
  {Gnedin}}]{2019MNRAS.487..364L}
{Li}, H., {Vogelsberger}, M., {Marinacci}, F., \& {Gnedin}, O.~Y. 2019, \mnras,
  487, 364

\bibitem[{{L{\"o}hner}(1987)}]{Lohner1987CMAME..61..323L}
{L{\"o}hner}, R. 1987, Computer Methods in Applied Mechanics and Engineering,
  61, 323

\bibitem[{{Mac Low}(1999)}]{mac-low1999a}
{Mac Low}, M.-M. 1999, \apj, 524, 169

\bibitem[{{McKee} \& {Ostriker}(2007)}]{2007ARA&A..45..565M}
{McKee}, C.~F. \& {Ostriker}, E.~C. 2007, \araa, 45, 565

\bibitem[{{McMillan} {et~al.}(2012){McMillan}, {Portegies Zwart}, {van
  Elteren}, \& {Whitehead}}]{mcmillan2012}
{McMillan}, S., {Portegies Zwart}, S., {van Elteren}, A., \& {Whitehead}, A.
  2012, in Astronomical Society of the Pacific Conference Series, Vol. 453,
  Advances in Computational Astrophysics: Methods, Tools, and Outcome, ed.
  R.~{Capuzzo-Dolcetta}, M.~{Limongi}, \& A.~{Tornamb{\`e}}, 129

\bibitem[{{Megeath} {et~al.}(2016){Megeath}, {Gutermuth}, {Muzerolle},
  {Kryukova}, {Hora}, {Allen}, {Flaherty}, {Hartmann}, {Myers}, {Pipher},
  {Stauffer}, {Young}, \& {Fazio}}]{2016AJ....151....5M}
{Megeath}, S.~T., {Gutermuth}, R., {Muzerolle}, J., {et~al.} 2016, \aj, 151, 5

\bibitem[{{Miyoshi} \& {Kusano}(2005)}]{2005JCoPh.208..315M}
{Miyoshi}, T. \& {Kusano}, K. 2005, Journal of Computational Physics, 208, 315

\bibitem[{{Motte} {et~al.}(2022){Motte}, {Bontemps}, {Csengeri}, {Pouteau},
  {Louvet}, {Stutz}, {Cunningham}, {L{\'o}pez-Sepulcre}, {Brouillet},
  {Galv{\'a}n-Madrid}, {Ginsburg}, {Maud}, {Men'shchikov}, {Nakamura}, {Nony},
  {Sanhueza}, {{\'A}lvarez-Guti{\'e}rrez}, {Armante}, {Baug}, {Bonfand},
  {Busquet}, {Chapillon}, {D{\'\i}az-Gonz{\'a}lez}, {Fern{\'a}ndez-L{\'o}pez},
  {Guzm{\'a}n}, {Herpin}, {Liu}, {Olguin}, {Towner}, {Bally}, {Battersby},
  {Braine}, {Bronfman}, {Chen}, {Dell'Ova}, {Di Francesco}, {Gonz{\'a}lez},
  {Gusdorf}, {Hennebelle}, {Izumi}, {Joncour}, {Lee}, {Lefloch}, {Lesaffre},
  {Lu}, {Menten}, {Mignon-Risse}, {Molet}, {Moraux}, {Mundy}, {Nguyen Luong},
  {Reyes}, {Reyes Reyes}, {Robitaille}, {Rosolowsky}, {Sandoval-Garrido},
  {Schuller}, {Svoboda}, {Tatematsu}, {Thomasson}, {Walker}, {Wu}, {Whitworth},
  \& {Wyrowski}}]{2022A&A...662A...8M}
{Motte}, F., {Bontemps}, S., {Csengeri}, T., {et~al.} 2022, \aap, 662, A8

\bibitem[{{Murray} {et~al.}(2010){Murray}, {Quataert}, \&
  {Thompson}}]{2010ApJ...709..191M}
{Murray}, N., {Quataert}, E., \& {Thompson}, T.~A. 2010, \apj, 709, 191

\bibitem[{{Myers}(2009)}]{hfs1-2009}
{Myers}, P.~C. 2009, \apj, 700, 1609

\bibitem[{{O'Connell}(2004)}]{2004ASPC..322..551O}
{O'Connell}, R.~W. 2004, in Astronomical Society of the Pacific Conference
  Series, Vol. 322, The Formation and Evolution of Massive Young Star Clusters,
  ed. H.~J.~G.~L.~M. {Lamers}, L.~J. {Smith}, \& A.~{Nota} (San Francisco:
  Astron.\ Soc.\ Pacific), 551

\bibitem[{{O'Dell}(2001)}]{2001ARA&A..39...99O}
{O'Dell}, C.~R. 2001, \araa, 39, 99

\bibitem[{{Padoan} \& {Nordlund}(2011)}]{2011ApJ...730...40P}
{Padoan}, P. \& {Nordlund}, {\r{A}}. 2011, \apj, 730, 40

\bibitem[{{Parmentier} \& {Pfalzner}(2013)}]{2013A&A...549A.132P}
{Parmentier}, G. \& {Pfalzner}, S. 2013, \aap, 549, A132

\bibitem[{{Pelupessy} \& {Portegies Zwart}(2012)}]{2012MNRAS.420.1503P}
{Pelupessy}, F.~I. \& {Portegies Zwart}, S. 2012, \mnras, 420, 1503

\bibitem[{{Pelupessy} {et~al.}(2013){Pelupessy}, {van Elteren}, {de Vries},
  {McMillan}, {Drost}, \& {Portegies Zwart}}]{2013A&A...557A..84P}
{Pelupessy}, F.~I., {van Elteren}, A., {de Vries}, N., {et~al.} 2013, \aap,
  557, A84

\bibitem[{Plummer(1911)}]{plummer1911}
Plummer, H.~C. 1911, Monthly Notices of the Royal Astronomical Society, 71, 460

\bibitem[{{Polak} {et~al.}(2025){Polak}, {Mac Low}, {Klessen}, {Portegies
  Zwart}, {Andersson}, {Appel}, {Cournoyer-Cloutier}, {Glover}, \&
  {McMillan}}]{polak2025}
{Polak}, B., {Mac Low}, M.-M., {Klessen}, R.~S., {et~al.} 2025, \aap, 695, A188

\bibitem[{{Polak} {et~al.}(2024){Polak}, {Mac Low}, {Klessen}, {Wei Teh},
  {Cournoyer-Cloutier}, {Andersson}, {Appel}, {Tran}, {Lewis}, {Wilhelm},
  {Portegies Zwart}, {Glover}, {Rieder}, {Wang}, \& {McMillan}}]{Polak_2024a}
{Polak}, B., {Mac Low}, M.-M., {Klessen}, R.~S., {et~al.} 2024, \aap, 690, A94

\bibitem[{{Portegies Zwart} \& {McMillan}(2018)}]{2018araa.book.....P}
{Portegies Zwart}, S. \& {McMillan}, S. 2018, {Astrophysical Recipes; The art
  of AMUSE} (IOP Publishing)

\bibitem[{{Portegies Zwart} {et~al.}(2009){Portegies Zwart}, {McMillan},
  {Harfst}, {Groen}, {Fujii}, {Nuall{\'a}in}, {Glebbeek}, {Heggie}, {Lombardi},
  {Hut}, {Angelou}, {Banerjee}, {Belkus}, {Fragos}, {Fregeau}, {Gaburov},
  {Izzard}, {Juri{\'c}}, {Justham}, {Sottoriva}, {Teuben}, {van Bever},
  {Yaron}, \& {Zemp}}]{2009NewA...14..369P}
{Portegies Zwart}, S., {McMillan}, S., {Harfst}, S., {et~al.} 2009, \na, 14,
  369

\bibitem[{{Portegies Zwart} {et~al.}(2013){Portegies Zwart}, {McMillan}, {van
  Elteren}, {Pelupessy}, \& {de Vries}}]{2013CoPhC.184..456P}
{Portegies Zwart}, S., {McMillan}, S.~L.~W., {van Elteren}, E., {Pelupessy},
  I., \& {de Vries}, N. 2013, Computer Physics Communications, 184, 456

\bibitem[{{Portegies Zwart} {et~al.}(2020){Portegies Zwart}, {Pelupessy},
  {Mart{\'\i}nez-Barbosa}, {van Elteren}, \& {McMillan}}]{2020CNSNS..8505240P}
{Portegies Zwart}, S., {Pelupessy}, I., {Mart{\'\i}nez-Barbosa}, C., {van
  Elteren}, A., \& {McMillan}, S. 2020, Communications in Nonlinear Science and
  Numerical Simulations, 85, 105240

\bibitem[{{Portegies Zwart} {et~al.}(2019){Portegies Zwart}, {van Elteren},
  {Pelupessy}, {McMillan}, {Rieder}, {de Vries}, {Marosvolgyi}, {Whitehead},
  {Wall}, {Drost}, {Jilkova}, {Martinez Barbosa}, {van der Helm}, {Beedorf},
  {Bos}, {Boekholt}, {van Werkhoven}, {Wijnen}, {Hamers}, {Caputo}, {Ferrari},
  {Toonen}, {Gaburov}, {Paardekooper}, {Janes}, {Punzo}, {Kruip}, \&
  {Altay}}]{2019zndo...3260650P}
{Portegies Zwart}, S., {van Elteren}, A., {Pelupessy}, I., {et~al.} 2019,
  {AMUSE: the Astrophysical Multipurpose Software Environment}

\bibitem[{{Portegies Zwart} \& {Verbunt}(1996)}]{portegies-zwart1996}
{Portegies Zwart}, S.~F. \& {Verbunt}, F. 1996, \aap, 309, 179

\bibitem[{{Proszkow} \& {Adams}(2009)}]{2009ApJS..185..486P}
{Proszkow}, E.-M. \& {Adams}, F.~C. 2009, \apjs, 185, 486

\bibitem[{{Rahner} {et~al.}(2019){Rahner}, {Pellegrini}, {Glover}, \&
  {Klessen}}]{Rahner+2019}
{Rahner}, D., {Pellegrini}, E.~W., {Glover}, S. C.~O., \& {Klessen}, R.~S.
  2019, \mnras, 483, 2547

\bibitem[{{Rhea} {et~al.}(2025){Rhea}, {Hlavacek-Larrondo},
  {Gendron-Marsolais}, {Vigneron}, {Donahue}, {Thilloy}, {Rousseau-Nepton},
  {Mezcua}, {Werner}, {Barrera-Ballesteros}, {Choi}, {Edge}, {Fabian}, \&
  {Voit}}]{2025AJ....169..203R}
{Rhea}, C.~L., {Hlavacek-Larrondo}, J., {Gendron-Marsolais}, M.-L., {et~al.}
  2025, \aj, 169, 203

\bibitem[{{Ricker}(2008)}]{ricker2008}
{Ricker}, P.~M. 2008, \apjs, 176, 293

\bibitem[{{Rogers} \& {Pittard}(2013)}]{2013MNRAS.431.1337R}
{Rogers}, H. \& {Pittard}, J.~M. 2013, \mnras, 431, 1337

\bibitem[{{Sales} {et~al.}(2014){Sales}, {Marinacci}, {Springel}, \&
  {Petkova}}]{Sales+2014}
{Sales}, L.~V., {Marinacci}, F., {Springel}, V., \& {Petkova}, M. 2014, \mnras,
  439, 2990

\bibitem[{{Schroetter} {et~al.}(2025){Schroetter}, {Berne}, {Boyden}, {Amiot},
  {Ballering}, {Canin}, {Cleeves}, {Goicoechea}, {Haworth}, {Joblin}, {Le
  Petit}, {Rogers}, \& {Sabbah}}]{2025jwst.prop.7534S}
{Schroetter}, I., {Berne}, O., {Boyden}, R., {et~al.} 2025, {A MIRI
  spectroscopic atlas of irradiated disks in Orion}, JWST Proposal. Cycle 4,
  ID. \#7534

\bibitem[{{Seifried} {et~al.}(2019){Seifried}, {Walch}, {Reissl}, \&
  {Ib{\'a}{\~n}ez-Mej{\'\i}a}}]{2019MNRAS.482.2697S}
{Seifried}, D., {Walch}, S., {Reissl}, S., \& {Ib{\'a}{\~n}ez-Mej{\'\i}a},
  J.~C. 2019, \mnras, 482, 2697

\bibitem[{{Shukirgaliyev}(2018)}]{BekPhDT}
{Shukirgaliyev}, B. 2018, PhD thesis, Astronomisches Rechen-Institut

\bibitem[{{Shukirgaliyev} {et~al.}(2019{\natexlab{a}}){Shukirgaliyev}, Otebay,
  Just, Berczik, Omarov, Naurzbaeva, \& Kalambay}]{shukirgaliyev2019violent}
{Shukirgaliyev}, B., Otebay, A., Just, A., {et~al.} 2019{\natexlab{a}},
  Proceedings of the National Academy of Sciences of the Republic of
  Kazakhstan. Physical and Mathematical Series, 130

\bibitem[{{Shukirgaliyev} {et~al.}(2021){Shukirgaliyev}, {Otebay}, {Sobolenko},
  {Ishchenko}, {Borodina}, {Panamarev}, {Myrzakul}, {Kalambay}, {Naurzbayeva},
  {Abdikamalov}, \& et~al.}]{Bek+2021}
{Shukirgaliyev}, B., {Otebay}, A., {Sobolenko}, M., {et~al.} 2021, \aap, 654,
  A53

\bibitem[{{Shukirgaliyev} {et~al.}(2017){Shukirgaliyev}, {Parmentier},
  {Berczik}, \& {Just}}]{Bek+2017}
{Shukirgaliyev}, B., {Parmentier}, G., {Berczik}, P., \& {Just}, A. 2017, \aap,
  605, A119

\bibitem[{{Shukirgaliyev} {et~al.}(2019{\natexlab{b}}){Shukirgaliyev},
  {Parmentier}, {Berczik}, \& {Just}}]{bek+2019}
{Shukirgaliyev}, B., {Parmentier}, G., {Berczik}, P., \& {Just}, A.
  2019{\natexlab{b}}, \mnras, 486, 1045

\bibitem[{{Shukirgaliyev} {et~al.}(2020){Shukirgaliyev}, {Parmentier},
  {Berczik}, \& {Just}}]{2020IAUS..351..507S}
{Shukirgaliyev}, B., {Parmentier}, G., {Berczik}, P., \& {Just}, A. 2020, in
  IAU Symposium, Vol. 351, Star Clusters: From the Milky Way to the Early
  Universe, ed. A.~{Bragaglia}, M.~{Davies}, A.~{Sills}, \& E.~{Vesperini}
  (Cambridge, UK: Cambridge U. Press), 507--511

\bibitem[{{Shukirgaliyev} {et~al.}(2018){Shukirgaliyev}, {Parmentier}, {Just},
  \& {Berczik}}]{bek+2018}
{Shukirgaliyev}, B., {Parmentier}, G., {Just}, A., \& {Berczik}, P. 2018, \apj,
  863, 171

\bibitem[{{Simpson} {et~al.}(2015){Simpson}, {Bryan}, {Hummels}, \&
  {Ostriker}}]{simpson2015}
{Simpson}, C.~M., {Bryan}, G.~L., {Hummels}, C., \& {Ostriker}, J.~P. 2015,
  \apj, 809, 69

\bibitem[{{Smith} {et~al.}(2011){Smith}, {Fellhauer}, {Goodwin}, \&
  {Assmann}}]{2011MNRAS.414.3036S}
{Smith}, R., {Fellhauer}, M., {Goodwin}, S., \& {Assmann}, P. 2011, \mnras,
  414, 3036

\bibitem[{{Sormani} {et~al.}(2017){Sormani}, {Tre{\ss}}, {Klessen}, \&
  {Glover}}]{2017MNRAS.466..407S}
{Sormani}, M.~C., {Tre{\ss}}, R.~G., {Klessen}, R.~S., \& {Glover}, S. C.~O.
  2017, \mnras, 466, 407

\bibitem[{{Springel}(2010)}]{2010MNRAS.401..791S}
{Springel}, V. 2010, \mnras, 401, 791

\bibitem[{{Stahler} \& {Palla}(2004)}]{stahler2004}
{Stahler}, S.~W. \& {Palla}, F. 2004, {The Formation of Stars} (Wiley), 865

\bibitem[{{Stutz}(2018)}]{2018MNRAS.473.4890S}
{Stutz}, A.~M. 2018, \mnras, 473, 4890

\bibitem[{{Toonen} {et~al.}(2012){Toonen}, {Nelemans}, \& {Portegies
  Zwart}}]{2012A&A...546A..70T}
{Toonen}, S., {Nelemans}, G., \& {Portegies Zwart}, S. 2012, \aap, 546, A70

\bibitem[{{Truelove} {et~al.}(1997){Truelove}, {Klein}, {McKee}, {Holliman},
  {Howell}, \& {Greenough}}]{1997ApJ...489L.179T}
{Truelove}, J.~K., {Klein}, R.~I., {McKee}, C.~F., {et~al.} 1997, \apjl, 489,
  L179

\bibitem[{Ussipov {et~al.}(2024)Ussipov, Akhmetali, Zaidyn, Akniyazova, Sakan,
  Mukhagali, \& Bekdaulet}]{ussipov2024fractal}
Ussipov, N., Akhmetali, A., Zaidyn, M., {et~al.} 2024, Eurasian Physical
  Technical Journal, 21, 108

\bibitem[{{Virtanen} {et~al.}(2020){Virtanen}, {Gommers}, {Oliphant},
  {Haberland}, {Reddy}, {Cournapeau}, {Burovski}, {Peterson}, {Weckesser},
  {Bright}, {van der Walt}, {Brett}, {Wilson}, {Millman}, {Mayorov}, {Nelson},
  {Jones}, {Kern}, {Larson}, {Carey}, {Polat}, {Feng}, {Moore}, {VanderPlas},
  {Laxalde}, {Perktold}, {Cimrman}, {Henriksen}, {Quintero}, {Harris},
  {Archibald}, {Ribeiro}, {Pedregosa}, {van Mulbregt}, \& {SciPy 1. 0
  Contributors}}]{2020NatMe..17..261V}
{Virtanen}, P., {Gommers}, R., {Oliphant}, T.~E., {et~al.} 2020, Nature
  Methods, 17, 261

\bibitem[{{Wall} {et~al.}(2020){Wall}, {Mac Low}, {McMillan}, {Klessen},
  {Portegies Zwart}, \& {Pellegrino}}]{wall2020}
{Wall}, J.~E., {Mac Low}, M.-M., {McMillan}, S. L.~W., {et~al.} 2020, \apj,
  904, 192

\bibitem[{{Wall} {et~al.}(2019){Wall}, {McMillan}, {Mac Low}, {Klessen}, \&
  {Portegies Zwart}}]{wall2019}
{Wall}, J.~E., {McMillan}, S. L.~W., {Mac Low}, M.-M., {Klessen}, R.~S., \&
  {Portegies Zwart}, S. 2019, \apj, 887, 62

\bibitem[{{Wise} \& {Abel}(2011)}]{wise2011}
{Wise}, J.~H. \& {Abel}, T. 2011, \mnras, 414, 3458

\bibitem[{{Wu} {et~al.}(2020){Wu}, {Tan}, {Christie}, \&
  {Nakamura}}]{2020ApJ...891..168W}
{Wu}, B., {Tan}, J.~C., {Christie}, D., \& {Nakamura}, F. 2020, \apj, 891, 168

\bibitem[{{W{\"u}nsch}(2015)}]{2015HiA....16..614W}
{W{\"u}nsch}, R. 2015, Highlights of Astronomy, 16, 614

\bibitem[{{Yan} {et~al.}(2023){Yan}, {Jerabkova}, \&
  {Kroupa}}]{2023A&A...670A.151Y}
{Yan}, Z., {Jerabkova}, T., \& {Kroupa}, P. 2023, \aap, 670, A151

\bibitem[{{Zhang} {et~al.}(2024{\natexlab{a}}){Zhang}, {Zhou}, {Esimbek},
  {Baan}, {He}, {Tang}, {Li}, {Ji}, {Wu}, {Ma}, {Li}, {Zhou}, {Tursun}, \&
  {Komesh}}]{hfs2-2024}
{Zhang}, W., {Zhou}, J., {Esimbek}, J., {et~al.} 2024{\natexlab{a}}, \aap, 688,
  A99

\bibitem[{{Zhang} {et~al.}(2024{\natexlab{b}}){Zhang}, {Zhou}, {Esimbek},
  {Baan}, {Tang}, {Li}, {He}, {Wu}, {Zhou}, {Ma}, {Tursun}, {Ji}, {Chang},
  {Li}, \& {Komesh}}]{hfs3-2024}
{Zhang}, W., {Zhou}, J., {Esimbek}, J., {et~al.} 2024{\natexlab{b}}, \apjs,
  275, 7

\end{thebibliography}
\bibliographystyle{aa}

\appendix
\onecolumn

\section{Density profile of unprocessed gas}
\label{app:density}

\citet{2013A&A...549A.132P} proposed a semi-analytical model in which a star cluster is formed from a centrally concentrated spherically symmetric gas cloud with constant SFE per free-fall time $\epsilon_{\text{ff}}$. The overall density profile $\rho_0(r)$ does not change, where $r$ is the distance from the center of the cluster. As stars form, the gas profile $\rho_\mathrm{gas}(r,t)$ and the star profile $\rho_\star(r,t)$ change depending on the evolution time $t$, but must satisfy the constraint
\begin{equation}
\rho_0(r) = \rho_{\text{gas}}(r,t) + \rho_{\star}(r,t).
\end{equation}
\citet{2013A&A...549A.132P} derive the gas and stellar profile evolution to be
\begin{equation}
\label{eq:rho_gas_app}
\rho_{\text{gas}}(t, r) = \left( \rho_0(r)^{-1/2} + \sqrt{\frac{8G}{3\pi}} \epsilon_{\text{ff}} t \right)^{-2},
\end{equation}
\begin{equation}
\label{eq:rho_star_app}
\rho_{\star}(t, r) = \rho_0(r) - \rho_{\rm gas}(r, t),
\end{equation}
where $G$ is the gravitational constant. \citet{Bek+2017} rewrote the constraint for a given density profile of the formed star cluster $\rho_{\star}(r)$,
\begin{equation}
\rho_0(r, t_{\text{SF}}) = \rho_{\text{gas}}(r, t_{\text{SF}}) + \rho_{\star}(r).
\end{equation}
In this formulation, the gas profile $\rho_{\text{gas}}$ and the total density profile $\rho_{0}$ depend on the star formation duration $t_{\text{SF}}$ and are calculated using Eqs. (\ref{eq:rho_gas_app}) and (\ref{eq:rho_star_app}), which results in 
\begin{equation}
    \rho_{\text{gas}} = \frac{1}{k^2} - \frac{\rho_{\star}}{2} - \frac{1}{2} \sqrt{K_2 + \frac{8}{k^6 K_1}} + K_1,
\end{equation}
where 
\begin{equation}
 K_0 = \frac{3 \sqrt{\alpha^3 + 36\alpha^2 + 216\alpha + 24 \alpha \sqrt{3(\alpha + 27)}}}{3(\alpha + 27)}, \quad \alpha = k^4 \rho_{\star}^2, \quad k = \sqrt{\frac{8G}{3\pi} } \epsilon_{\text{ff}} t_{\text{SF}}, \nonumber
\end{equation}
and 
\begin{equation}
K_1 = \frac{\sqrt{\alpha^2 + \alpha (K_0 + 24) + K_0 (K_0 + 12)}}{12 k^4 K_0}, \quad K_2 = \frac{(\alpha - K_0 + 24)(K_0 - \alpha)}{3 k^4 K_0}. \nonumber
\end{equation}

\end{document}